\documentclass[a4paper,english]{aa}
\usepackage{amsmath}
\usepackage{color}
\usepackage{graphicx}
\usepackage{amssymb}
\usepackage[]{natbib}

\def \kms{{\rm \, km \, {s}^{-1}}}

\makeatletter

\providecommand{\tabularnewline}{\\}

\authorrunning{Nehm\'e et al.}
\titlerunning{Multi-wavelength observations of a multi-phase cloud}

\usepackage{babel}
\makeatother
\begin{document}
\title{Multi-wavelength observations \\
 of a nearby  multi-phase  interstellar cloud }

\author{Cyrine Nehm\'e \inst{1}, C\'ecile Gry\inst{2,3},
Fran\c{c}ois Boulanger \inst{4},  Jacques Le Bourlot\inst{1},\\
Guillaume Pineau des For\^ets\inst{4,5}\and Edith Falgarone\inst{5}}

\institute{LUTH, Observatoire de Paris-Meudon, Universit\'e Paris 7, France \and 
Laboratoire d'Astrophysique de Marseille,OAMP, BP 8, 13376 Marseille Cedex 12, France \and
European Space Astronomy Center, RSSD, P O Box 50727, 28080 Madrid, Spain \and 
Institut d'Astrophysique Spatiale, Universit\'e Paris Sud, Bat. 121, 91405 Orsay Cedex, France
\and Laboratoire de Radio-Astronomie, LERMA, Ecole Normale Sup\'erieure, 24 rue
Lhomond, 75231 Paris Cedex 05, France }

\mail{cecile.gry@oamp.fr}

\date{Received / Accepted}
\abstract
{}
{High-resolution 
spectroscopic observations (UV HST/STIS and optical) are used to characterize the 
physical state and velocity structure of the multiphase interstellar medium seen
towards the nearby ($170\, $pc) star  HD~102065. The star is 
located behind the tail of a cometary-shaped, infrared cirrus-cloud,
in the area of interaction between the Sco-Cen OB association and the Local Bubble. }
{We analyze interstellar components present along the
line of sight
by fitting multiple transitions from a group of species all at once.
We identify four groups of species: (1) molecules (CO, CH, CH$^+$), (2) atoms (C~{\sc i}, S~{\sc i}, Fe~{\sc i}) 
with ionization potential lower than H~{\sc i}, (3)
neutral and low-ionized states of atoms (Mg~{\sc i}, Mg~{\sc ii}, Mn~{\sc ii}, P~{\sc ii}, 
Ni~{\sc ii}, C~{\sc ii}, N~{\sc i} and O~{\sc i}) 
with ionization potential larger than H~{\sc i} and (4) highly-ionized atoms (Si~{\sc iii},
C~{\sc iv}, Si~{\sc iv}). 
The absorption spectra are complemented by H~{\sc i}, CO and C~{\sc ii} emission-line spectra,
 H$_2$ column-densities derived
from FUSE spectra, and IRAS images.}
{Gas components 
of a wide range of temperatures and ionization states are detected along the line of sight.
Most of the hydrogen column-density is in cold, diffuse, molecular gas at 
low LSR velocity. 
This  gas is mixed with traces of 
warmer molecular gas traced by H$_2$ in the $J>2$ levels, in which   
the observed CH$^+$ must be formed.

We also identify three distinct components 
of warm gas at negative velocities down to $ -20\kms$. The temperature
and gas excitation are shown to increase with increasing velocity shift from the bulk of the gas. 

{        Hot gas  at   temperatures of several $10^5$ K  is detected in the most negative 
velocity component in the highly-ionized specie. This hot gas is also detected in very strong lines 
of less-ionized species (Mg~{\sc ii}, Si~{\sc ii}$^*$ and C~{\sc ii}$^*$) for which the bulk of the gas 
is cooler.}
}
{We relate the observational results to evidence for dynamical impact  of
the Sco-Cen stellar association on the nearby  interstellar medium. 
We propose a scenario where the infrared cirrus cloud 
has been hit a few $10^5\, $yr ago by a supernova blast wave originating 
from the Lower Centaurus Crux group of the Sco-Cen association.

The observations provide detailed information on the interplay between ISM phases in relation with
the origin of the Local and Loop I bubbles.}
\maketitle

\section{Introduction}

Astronomical observations provide multiple perspectives on the 
interstellar matter, gas and dust, 
that are rarely gathered together to study the relation between interstellar-medium phases.
It is the ambition of this paper to present such a study on gas and dust observed in the direction of the 
nearby (170 pc), lightly reddened (E(B-V) = 0.17), B9IV star HD~102065 in Chamaeleon (Table 1).  
In IRAS images, the infrared cirrus (Dcld 300.2-16.9) seen in the foreground  of
HD~102065 has a prominent cometary shape and an unusually blue IRAS colors (high 12 and $25\, \mu$m 
to $100\, \mu$m brightness ratio), 
indicative of a large abundance of small stochastically-heated 
dust particles (Boulanger et al. 1990).
We hereafter refer to it as the Blue Cloud in reference to its blue infrared colors. 
While the main Chamaeleon clouds
have a comparable extent in CO surveys as in IRAS images, only the center of the Blue Cloud 
head is detected in CO \citep{BBTD98,Mizuno2001}. Boulanger et al.  estimate that only 10\% 
of the cloud mass is seen in CO emission. On larger angular scales, 
HD~102065 lies within the sky area where various evidences for an
interaction between the Scorpius-Centaurus (Sco-Cen) OB association and the Local Bubble
have been gathered from H~{\sc i} observations (de Geus 1992), the X-ray ROSAT survey 
(Egger and Aschenbach 1995) and optical absorption-line spectroscopy (Corradi et al. 2004).

HD~102065 is located at the same Galactic longitude and slightly lower Galactic 
latitude than the Lower Centaurus Crux (LCC) group of the Sco-Cen association. 
Hipparcos positions, proper motions, and parallaxes have been used to show that 
the LCC and the Upper Centaurus Lupus (UCL) group of stars, presently located within the Loop I 
bubble at a distance 
of $\sim 120\, $pc, were located closer to the Sun $5-7\,$Myr ago. In 
two independent works (Maiz-Apellaniz 2001 and Bergh{\"o}fer and Breitschwerdt 2002), 
this information was used to 
relate the origin of the Local Bubble to the $\sim20$ supernovae explosions 
associated with the LCC and UCL groups that are estimated to have occurred 
in the last 10~Myr. 
The structure and physical state of the interstellar medium seen in direction of the
Sco-Cen association should reflect
the action of the supernovae explosions powering and 
progressively shredding, and spreading away the stars' parent cloud.
This scenario has been introduced into a numerical simulation based on the 
supernovae-driven ISM model of de Avillez (2000) by Breitschwerdt and de Avillez (2006).
In the reproduction of the local interstellar medium generated by this 
simulation (Figure 1 in Breitschwerdt and de Avillez), HD~102065 is located within the
Loop I bubble and the line of sight to the star crosses the shell that provides separation
from the Local Bubble.

Observations characterizing the
gas and the dust along the line of sight to HD102065 have been presented 
in earlier publications (Boulanger, Prevot \& Gry, 1994 ; Gry et al, 1998). 
Unlike most early-type stars observed in UV spectroscopy, the position of HD~102065 corresponds to no enhancement
 in IRAS emission,
implying that the matter responsible for absorption is not heated by the star \citep{BOUPREG}.
The extinction curve produced using IUE low-resolution
spectroscopy,  shows a strong 220~nm 
bump and a weak far-UV rise \citep{BOUPREG}. 
Moderate-resolution spectroscopy ($\lambda/\Delta\lambda=20\,000$) acquired by FUSE, detected 
molecular hydrogen absorption lines. These observations showed that
${\rm {H}}_{2}$ at a temperature of $\sim\,60\,$K is the dominant
state of hydrogen along the line of sight to HD~102065.
The presence of a smaller amount of warmer 
${\rm {H}_{2}}$ (a few 100 K) was inferred from the H$_2$ excitation
\citep{GBNPHF}.
HD~102065 was also observed using the Goddard High
Resolution Spectrometer (GHRS) on the Hubble Space Telescope (HST). These
observations revealed the presence of high, negative-velocity gas with 
unusually high ${\rm {Si~{\sc ii}}}$ excitation that was seen as the signature 
of the dissipation of a large amount of kinetic energy \citep{GBFPL}.
\begin{table}[!htbp]
\begin{center}\begin{tabular}{c|c}
\hline $\:$& HD~102065\tabularnewline 
\hline \hline $\alpha$ (2000)&11 43 37.87\tabularnewline 
$\delta$ (2000)& -80 28 59.4\tabularnewline
{\it l} (2000)&300.02$^\circ$\tabularnewline
{\it b} (2000)&-18.00$^\circ$\tabularnewline
Sp. Type& B9IV\tabularnewline 
d (pc)& 170\tabularnewline 
E(B-V)&$0.17\pm0.04$\tabularnewline
 $A_{v}$& $0.67\pm0.12$\tabularnewline
$R_{v}$& $3.9\pm0.4$\tabularnewline
N(${\rm{H}_{2}}$,J=0) (${\rm {cm}^{-2}}$)&$2.0\pm0.2\:10^{20}$\tabularnewline 
N(${\rm {H}_{2}}$,J=1) (${\rm{cm}^{-2}}$)& $1.4\pm0.1\:10^{20}$\tabularnewline 
N(${\rm{H}}_{2}$,J=2) (${\rm {cm}}^{-2}$)&$2.5-2.6\,10^{18}$\tabularnewline 
N(${\rm {H}_{2}}$,J=3) (${\rm{cm}^{-2}}$)& $1.1-3.1\,10^{17}$\tabularnewline 
N(${\rm{H}_{2}}$,J=4)$\,$(${\rm {cm}^{-2}}$)& $0.6-5.6\,10^{16}$\tabularnewline 
N(${\rm {H}}_{2}$,J=5) (${\rm{cm}}^{-2}$)& $0.45-1.2\,10^{15}$\tabularnewline 
N$_{H}$ (${\rm{cm}}^{-2}$)& $9.9\:10^{20}$\tabularnewline
$f=2N(H_{2}) / N_{H}$& $0.69\pm0.12$\tabularnewline
\hline
\end{tabular}\end{center}
\caption{\label{Tab_1} HD~102065 ``fact sheet''. $A_{v}$ is the
visible extinction, E(B-V) is the color index
\citep{BOUPREG}. ${\rm {H}_{2}}$ column densities from
\citep{GBNPHF}. The total column density (N$_{H}$ = N(H) + 2
N(H$_2$)) has been derived from E(B-V). 
}
\end{table}

A number of questions remain unanswered. Why is the small dust abundance enhanced 
in the Blue Cloud, which, from the FUSE and optical spectroscopy, appears to be 
a {\it typical} diffuse molecular cloud? How was the negative-velocity gas accelerated?  
Are the low and negative-velocity gas physically connected? More generally, how can the
data be interpreted within the present understanding of 
the interaction between the Loop~I and 
the Local bubbles? 

In this paper, we present high-resolution ($R=100\,000$) UV spectr,a
obtained using the HST Imaging Spectrograph  
(STIS) complemented by optical absorption spectra, 
and CO and H~{\sc i} emission spectroscopy (Sections~\ref{sec:UV-spectra} and 3). 
These data improve previous observations in spectral resolution and spectral coverage.
The multi-wavelength spectra are used to describe the physical state and velocity structure of 
the multiple gas components observed along the line of sight (Section~\ref{sec:data-analysis}). 
In Section~\ref{sec:model}, we discuss the cool, 
low-velocity gas, and 
in Section~\ref{sec:context}, 
we relate the observational results 
to evidence for dynamical interaction between the Sco-Cen stars and the nearby 
interstellar medium, and discuss the possible interplay between the multiple components.   
The main conclusions of this paper are gathered in Section~\ref{sec:Conclusions}.
\begin{center}%
\begin{table*}[t]
\begin{center}\begin{tabular}{lllccccclll}
\hline
{ Element}&{ Wavelength ($\textrm{\AA}$)}&{ f-value}&
&&&&&{ Element }&{ Wavelength ($\textrm{\AA}$)}&{ f-value}\tabularnewline
\hline
\hline
{\small ${\rm {C~{\sc I}}}$}&{\small 1276.4822}&{\small 0.449~$10^{-2}$}&
&&&&&{\small ${\rm {Fe~{\sc II}}}$}&{\small 1608.4510}&{\small 0.58~$10^{-1}$}\tabularnewline
{\small ${\rm {C~{\sc I}}}$}&{\small 1280.1353}&{\small 0.229~$10^{-1}$}&
&&&&&{\small ${\rm {Fe~{\sc II}}}$}&{\small 1611.2004}&{\small 0.136~$10^{-2}$}\tabularnewline
{\small ${\rm {C~{\sc I}}}$}&{1328.8333}&{0.631~$10^{-1}$ }&
&&&&&{\small ${\rm {Fe~{\sc II}}}$}&{\small 2586.6499}&{\small 0.691~$10^{-1}$}\tabularnewline
{\small ${\rm {C~{\sc I}}}$}&{\small 1560.3092}&{\small 0.128~$10^{-1}$}&
&&&&&{\small ${\rm {Fe~{\sc II}}}$}&{\small 2382.7651}&{\small 0.320}\tabularnewline
{\small ${\rm {C~{\sc I}}}$}&{\small 1656.9283}&{\small 0.140~$10^{0}$}&
&&&&&{\small ${\rm {Fe~{\sc II}}}$}&{\small 2600.1729}&{\small 0.239}\tabularnewline
{\small ${\rm {C~{\sc I}}^{*}}$}&{\small 1260.9261}&{\small 0.135~$10^{-1}$}&
&&&&&{\small ${\rm {Mn~{\sc II}}}$}&{\small 2576.8770}&{\small 0.361}\tabularnewline
{\small ${\rm {C~{\sc I}}^{*}}$}&{\small 1260.9961}&{\small 0.105~$10^{-1}$}&
&&&&&{\small ${\rm {Mn~{\sc II}}}$}&{\small 2594.4990}&{\small 0.280}\tabularnewline
{\small ${\rm {C~{\sc I}}^{*}}$}&{\small 1261.1224}&{\small 0.147~$10^{-1}$}&
&&&&&{\small ${\rm {Mn~{\sc II}}}$}&{\small 2606.4619}&{\small 0.198}\tabularnewline
{\small ${\rm {C~{\sc I}}^{*}}$}&{\small 1277.2827}&{\small 0.705~$10^{-1}$}&
&&&&&{\small ${\rm {Mg~{\sc II}}}$}&{\small 1239.9253}&{\small 0.617~$10^{-3}$}\tabularnewline
{\small ${\rm {C~{\sc I}}^{*}}$}&{\small 1277.5131}&{\small 0.223~$10^{-1}$}&
&&&&&{\small ${\rm {Mg~{\sc II}}}$}&{\small 1240.3947}&{\small 0.354~$10^{-3}$}\tabularnewline
{\small ${\rm {C~{\sc I}}^{*}}$}&{\small 1280.5975}&{\small 0.674~$10^{-2}$}&
&&&&&{\small ${\rm {Mg~{\sc II}}}$}&{\small 2803.5305}&{\small 0.306}\tabularnewline
{\small ${\rm {C~{\sc I}}^{*}}$}&{\small 1279.8907}&{\small 0.126~$10^{-1}$}&
&&&&&{\small ${\rm {Mg~{\sc II}}}$}&{\small 2796.3518}&{\small 0.615}\tabularnewline
{\small ${\rm {C~{\sc I}}^{*}}$}&{\small 1656.2672}&{\small 0.558~$10^{-1}$}&
&&&&&{\small ${\rm {S~{\sc II}}}$}&{\small 1250.5840}&{\small 0.543~$10^{-2}$}\tabularnewline
{\small ${\rm {C~{\sc I}}^{*}}$}&{\small 1657.3792}&{\small 0.356~$10^{-1}$}&
&&&&&{\small ${\rm {S~{\sc II}}}$}&{\small 1253.8110}&{\small 0.109~$10^{-1}$}\tabularnewline
{\small ${\rm {C~{\sc I}}^{*}}$}&{\small 1657.9071}&{\small 0.471~$10^{-1}$}&
&&&&&{\small ${\rm {S~{\sc II}}}$}&{\small 1259.5190}&{\small 0.166~$10^{-1}$}\tabularnewline
{\small ${\rm {C~{\sc I}}}^{**}$}&{\small 1277.5500}&{\small 0.791~$10^{-1}$}&
&&&&&{\small ${\rm {P~{\sc II}}}$}&{\small 1301.8743}&{\small 0.127~$10^{-1}$}\tabularnewline
{\small ${\rm {C~{\sc I}}^{**}}$}&{\small 1329.5775}&{\small 0.474~$10^{-1}$}&
&&&&&{\small ${\rm {P~{\sc II}}}$}&{\small 1532.5330}&{\small 0.303~$10^{-2}$}\tabularnewline
{\small ${\rm {C~{\sc I}}^{**}}$}&{\small 1329.6003}&{\small 0.159~$10^{-1}$}&
&&&&&{\small ${\rm {C~{\sc I}I}}$}&{\small 1334.5323}&{\small 0.128}\tabularnewline
{\small ${\rm {C~{\sc I}}^{**}}$}&{\small 1561.4377}&{\small 0.107~$10^{-1}$}&
&&&&&{\small ${\rm {C~{\sc II}}^{*}}$}&{\small 1335.7080}&{\small 0.115}\tabularnewline
{\small ${\rm {C~{\sc I}}^{**}}$}&{\small 1657.0081}&{\small 0.104~$10^{-1}$}&
&&&&&{\small ${\rm {\rm {Si~{\sc II}}}}$}&{\small 1304.3702}&{\small 0.917~$10^{-1}$}\tabularnewline
{\small ${\rm {S~{\sc I}}}$}&{\small 1295.6531}&{\small 0.870~$10^{-1}$}&
&&&&&{\small ${\rm {Si~{\sc II}}}$}&{\small 1193.2897}&{\small 0.585}\tabularnewline
{\small ${\rm {S~{\sc I}}}$}&{\small 1296.1740}&{\small 0.220~$10^{-1}$}&
&&&&&{\small ${\rm {Si~{\sc II}}}$}&{\small 1260.4221}&{\small 0.118~$10^{1}$}\tabularnewline
{\small ${\rm {S~{\sc I}}}$}&{\small 1425.1877}&{\small 0.365~$10^{-1}$}&
&&&&&{\small ${\rm {\rm {Si~{\sc II}}}}$}&{\small 1190.4158}&{\small 0.293}\tabularnewline
{\small ${\rm {S~{\sc I}}}$}&{\small 1473.9943}&{\small 0.730~$10^{-1}$}&
&&&&&{\small ${\rm {Si~{\sc II}}^{*}}$}&{\small 1264.7377}&{\small 0.106~$10^{1}$}\tabularnewline
{\small ${\rm {S~{\sc I}}}$}&{\small 1474.5706}&{\small 0.121~$10^{-2}$}&
&&&&&{\small ${\rm {Si~{\sc II}}^{*}}$}&{\small 1194.5002}&{\small 0.730}\tabularnewline
{\small ${\rm {Fe~{\sc I}}}$}&{\small 2523.6084}&{\small 0.279}&
&&&&&{\small ${\rm {Si~{\sc II}}^{*}}$}&{\small 1197.3938}&{\small 0.146}\tabularnewline
{\small ${\rm {Fe~{\sc I}}}$}&{\small 2484.0210}&{\small 0.557}&
&&&&&{\small ${\rm {Si~{\sc II}}^{*}}$}&{\small 1265.0020}&{\small 0.118}\tabularnewline
{\small${\rm {Mg~{\sc I}}}$}&{\small 2852.9641}&{\small 0.183~$10^{1}$}&
&&&&&{\small ${\rm {Si~{\sc II}}^{*}}$}&{\small 1309.2758}&{\small 0.913~$10^{-1}$}\tabularnewline
CO A-X (2,0) :&&&
&&&&&{\small ${\rm {Si~{\sc II}}}^{*}$}&{\small 1533.4321}&{\small 0.131}\tabularnewline
{\small ${\rm {R(0)}}$}&{\small 1477.5669}&{\small 3.933~$10^{-2}$}&
&&&&&{\small ${\rm {O~{\sc I}}}$}&{\small 1302.1685}&{\small 0.519~$10^{-1}$}\tabularnewline
{\small ${\rm {R(1)}}$}&{\small 1477.5148}&{\small 1.967~$10^{-2}$}&
&&&&&{\small ${\rm {O~{\sc I}}}$}&{\small 1355.5977}&{\small 0.116~$10^{-5}$}\tabularnewline
{\small ${\rm {R(2)}}$}&{\small 1477.4786}&{\small 1.574~$10^{-2}$}&
&&&&&{\small ${\rm {O~{\sc I}}^{*}}$}&{\small 1304.8576}&{\small 0.518~$10^{-1}$}\tabularnewline
{\small ${\rm {Q(1)}}$}&{\small 1477.6509}&{\small 1.967~$10^{-2}$}&
&&&&&{\small ${\rm {O~{\sc I}}}^{**}$}&{\small 1306.0286}&{\small 0.518~$10^{-1}$}\tabularnewline
{\small ${\rm {Q(2)}}$}&{\small 1477.6827}&{\small 1.967~$10^{-2}$}&
&&&&&{\small ${\rm {N~{\sc I}}}$}&{\small 1199.5496}&{\small 0.130}\tabularnewline
CO A-X (1,0) :&&&
&&&&&{\small ${\rm {N~{\sc I}}}$}&{\small 1200.2233}&{\small 0.862~$10^{-1}$}\tabularnewline
{\small ${\rm {R(0)}}$}&{\small 1509.7504}&{\small 2.855~$10^{-2}$}&
&&&&&{\small ${\rm {N~{\sc I}}}$}&{\small 1200.7098}&{\small 0.430~$10^{-1}$}\tabularnewline
{\small ${\rm {R(1)}}$}&{\small 1509.6985}&{\small 1.449~$10^{-2}$}&
&&&&&\textrm{\small ${\rm {N~{\sc I}}}$}&\textrm{\small 1159.8168}&\textrm{\small 0.851~$10^{-5}$}\tabularnewline
{\small ${\rm {\rm {R(2)}}}$}&{\small 1509.6639}&{\small 1.183~$10^{-2}$}&
&&&&&\textrm{\small ${\rm {N~{\sc I}}}$}&\textrm{\small 1160.9366}&\textrm{\small 0.240~$10^{-5}$}\tabularnewline
{\small ${\rm {\rm {Q(1)}}}$}&{\small 1509.8379}&{\small 1.427~$10^{-2}$}&
&&&&&{\small ${\rm {Ni~{\sc II}}}$}&{\small 1317.2170}&{\small 0.775~$10^{-1}$}\tabularnewline
{\small ${\rm {Q(2)}}$}&{\small 1509.8738}&{\small 1.449~$10^{-2}$}&
&&&&&{\small ${\rm {Ni~{\sc II}}}$}&{\small 1370.1320}&{\small 0.131}\tabularnewline
CO A-X (0,0) :&&&
&&&&&\textrm{\small ${\rm {Ni~{\sc II}}}$}&\textrm{\small 1454.8420}&\textrm{\small 0.595~$10^{-1}$}\tabularnewline
{\small ${\rm {R(0)}}$}&{\small 1544.4515}&{\small 1.598~$10^{-2}$}&
&&&&&\textrm{\small ${\rm {C~{\sc IV}}}$}&\textrm{\small 1550.7770}&\textrm{\small 0.948~$10^{-1}$}\tabularnewline
{\small ${\rm {R(1)}}$}&{\small 1544.3914}&{\small 7.977~$10^{-3}$}&
&&&&&\textrm{\small ${\rm {C~{\sc IV}}}$}&\textrm{\small 1548.2030}&\textrm{\small 0.190}\tabularnewline
{\small ${\rm {R(2)}}$}&{\small 1544.3472}&{\small 6.361~$10^{-3}$}&
&&&&&\textrm{\small ${\rm {Si~{\sc III}}}$}&\textrm{\small 1206.500}&\textrm{\small 1.67}\tabularnewline
{\small ${\rm {Q(1)}}$}&{\small 1544.5432}&{\small 7.993~$10^{-3}$}&
&&&&&\textrm{\small ${\rm {Si~{\sc IV}}}$}&\textrm{\small 1393.755}&\textrm{\small 0.514}\tabularnewline
{\small ${\rm {Q(2)}}$}&{\small 1544.5743}&{\small 7.983~$10^{-3}$}&
&&&&&\textrm{\small ${\rm {Si~{\sc IV}}}$}&\textrm{\small 1402.770}&\textrm{\small 0.255}\tabularnewline
\hline
\end{tabular}\end{center}
\caption{Wavelength and $f$-values for the observed absorption lines. Most
of those values are from Morton (2000 and 2001) and Verner et al. (1994).
Oscillator strengths values for S~{\sc i} are taken from Beideck et al. (1994). Values for
O~{\sc i} $1355.5977\,\textrm{\AA}$  and Ni~{\sc ii} $1317.2170\,\textrm{\AA}$
are from Welty et al. (1999). CO oscillator strengths have been kindly provided by Dr. Eidelsberg. \label{fvalues}}
\end{table*}
\end{center}
\section{Ultraviolet absorption lines: STIS data. \label{sec:UV-spectra}}
\subsection{Observations}
We acquired STIS UV spectra of HD~102065 with
the high-resolution MAMA Echelle gratings E140H and E230H. We used
five different grating settings covering the following wavelength
regions: 1144 - 1324 \AA, 1322 - 1512 \AA, and 1494 - 1674 \AA\
(E140H with aperture $0.2\times0.09)$, and 2376 - 2640 \AA\ and 2620
- 2864 \AA\ (E230H with aperture $0.1\times0.09$).

We performed  spectral extraction using the calibration software
CALSTIS version 2.14c. This version provided robust calibration
of the zero-flux level,  an important parameter when deriving
column densities from absorption lines: in all strongly saturated
lines, the mean zero-level is 4 to 5 times below the RMS noise. We
therefore assume that the zero-flux level is valid for all
spectra, and we do not consider it a free parameter when fitting the
absorption profiles.
\begin{figure}[!htbp]
\begin{center}\includegraphics[%
  width=1.\columnwidth]{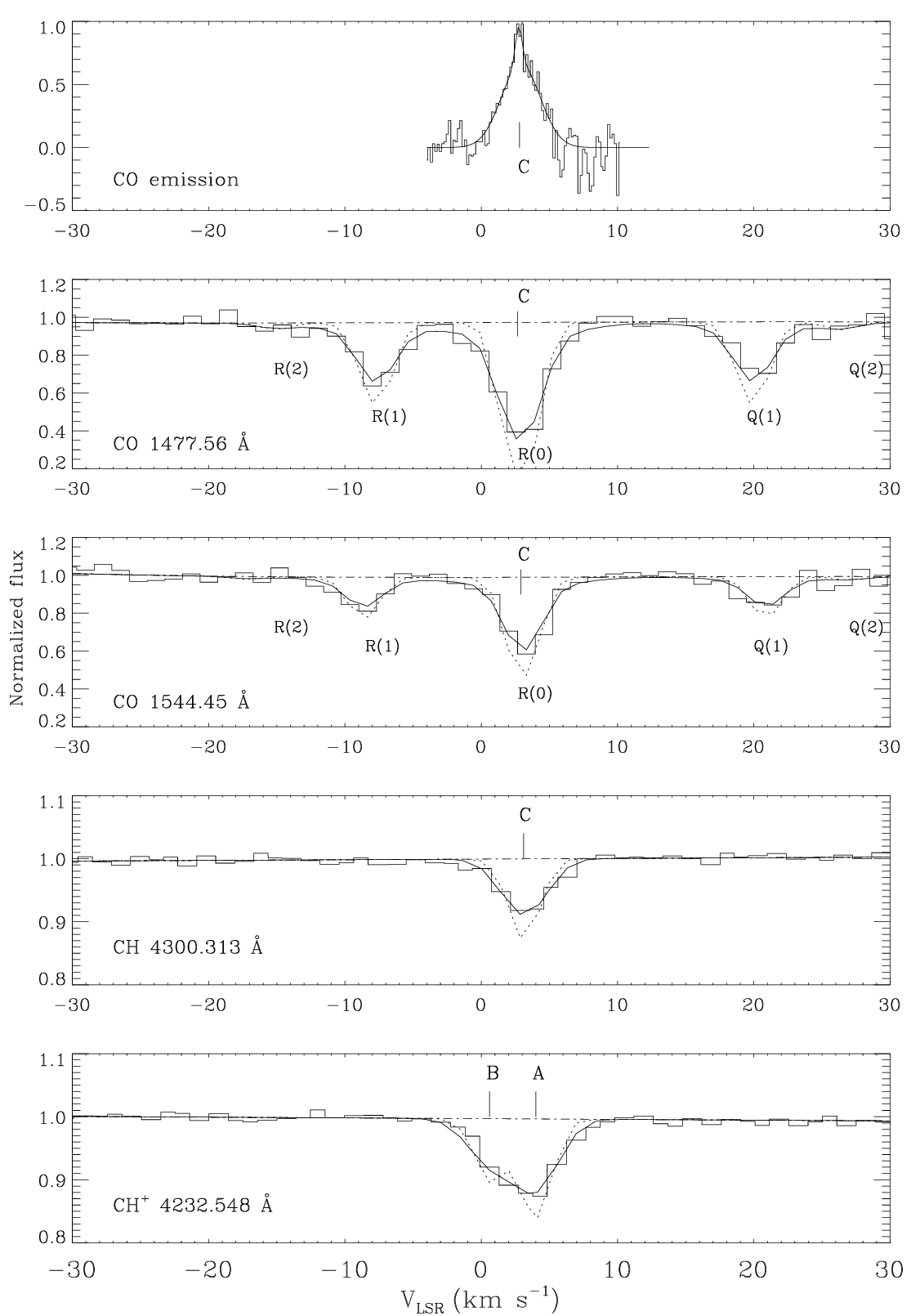}
\end{center}
\caption{Molecular lines. From top to bottom: $^{12}$CO J=2-1 emission line (Section 3.3) 
in the millimeter range with the SEST, two examples of the CO ultraviolet
absorption bands (A-X (2,0) and A-X (0,0)) observed with STIS, and the CH and CH$^+$ visible absorption lines (Section 3.1). 
Histogram-style curves represent the observations normalized to the continuum,
solid lines are the best fits, and dashed lines, the profiles before convolution
with the instrument line spread function, i.e. the intrinsic interstellar profile. 
The letters A, B and C indicate
velocity components that will be introduced in the spectral analysis. 
\label{fig:molecules}}
\end{figure}
\begin{figure}[!htbp]
\includegraphics[width=1.\columnwidth]{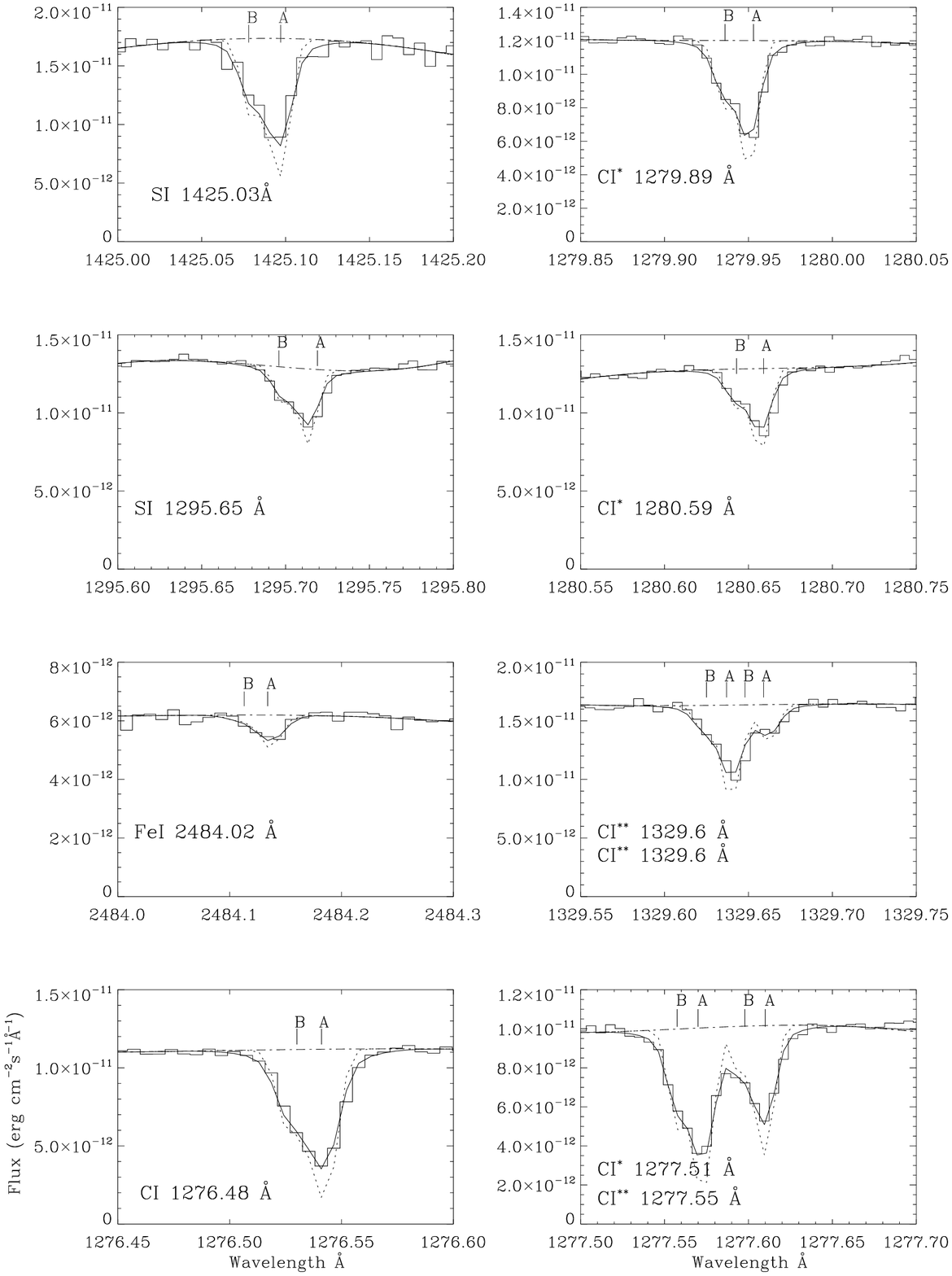}
\caption{Sample lines from neutral species.  Histogram-style curves represent the observations 
(in ${\rm erg\, cm^{-2}\, s^{-1}}\textrm{Å}^{-1}$),
solid lines are the best fits, and dashed lines, the profiles before convolution
with the instrument line spread function, i.e. the intrinsic interstellar profile. An asymmetry is evident on most profiles 
which are therefore best fitted with two components, called A and B.
\label{fig:neutrals}}
\end{figure}

For the wavelength region 1144 - 1324 \AA ,  observations
were completed in five sub-exposures. These five measurements do not have
the same wavelength grid and for each wavelength step the five wavelength
values are distributed over the entire resolution element. In order to
not  degrade the resolution when co-adding the five sub-exposures,
we have defined a mean wavelength grid and have interpolated each
sub-exposure spectrum over this grid in common. The resulting smoothing
factor with this method is no more than 15\%.

The resolution achieved with the echelle gratings is about $2.6 \kms$.
The line spread functions (LSF) are tabulated in the STIS Instrument
Handbook.  We have performed a  double-Gaussian fit 
to the tabulated LSF and use the resulting FWHM values in the profile-fitting software.
 The results of the double-Gaussian fit to the LSF
were a narrow component with FWHM $\thicksim$ 1.2 pixels (6.3 m\AA ) for E140H
at 1200 \AA, 1.1 pixels (7.2 m\AA) for E140H at 1500 \AA, and 1.7 pixels (17.9m\AA) for
 E230H at 2400 \AA, and a broad component with FWHM $\thicksim$
6 pixels in all cases.

According to the STIS Instrument Handbook, the absolute wavelength
accuracy across exposures is 0.2-0.5 pixels. We verified that these measurements were applicable to our own observations
by comparing the centers of spectral lines corresponding to the same
element, occurring in different exposures at different wavelengths:
shifts between different wavelength ranges were never larger than 0.5
pixel, or $0.6 \kms$.

The achieved signal-to-noise ratio (hereafter SNR) in the continuum, ranges from 23
to 41 depending on the grating setting. At the positions
of most interstellar lines the SNR ranges from 20 to 40, except
for the lines Si~{\sc ii}*, C~{\sc ii} and C~{\sc ii}*,
where it drops to 6 because they are located at the bottom of strong
stellar lines.
\subsection{A guided tour of the STIS absorption spectra}
{        Here we provide an overview of  the   
 interstellar, absorption lines present in the STIS spectra and  used in the analysis. 
 They are listed in Table~\ref{fvalues}.}

\begin{figure*}[!htbp]
\begin{center}\includegraphics[%
height=1.0\paperwidth,
keepaspectratio]{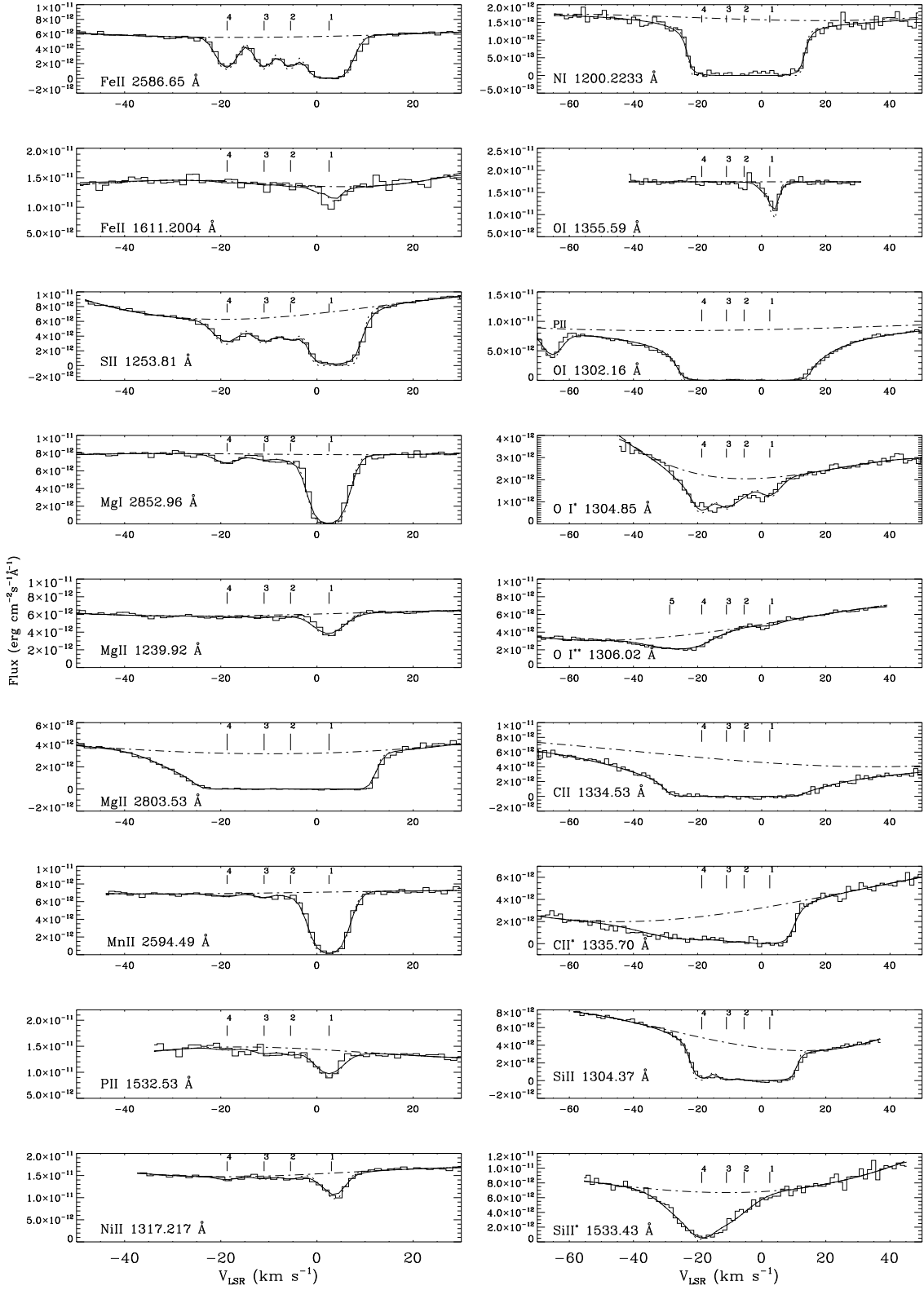}
\end{center}
\caption{Sample of lines from ionized species and neutral 
species with ionization
potential close to that of H~{\sc i}. 
Same line style as Figure~\ref{fig:neutrals}.
Most fits have been performed with the same line-of-sight model, made of 4 components with same 
velocity and same temperature for
all species. There are only three exceptions: (i) An additional component of high-temperature gas is 
included at the velocity of component~4 to fit the  extended blue wing of the strong Mg~{\sc ii}
and C~{\sc ii}* lines. (ii) In the O~{\sc i}** profile an additional component is 
considered, blueward of component~4. (iii) The S~{\sc ii} profiles include 
an extra absorption on the red side of component 1.\label{fig:ions}}
\end{figure*}

\begin{figure}[!htbp]
\begin{center}\includegraphics[%
  width=1.0\columnwidth]{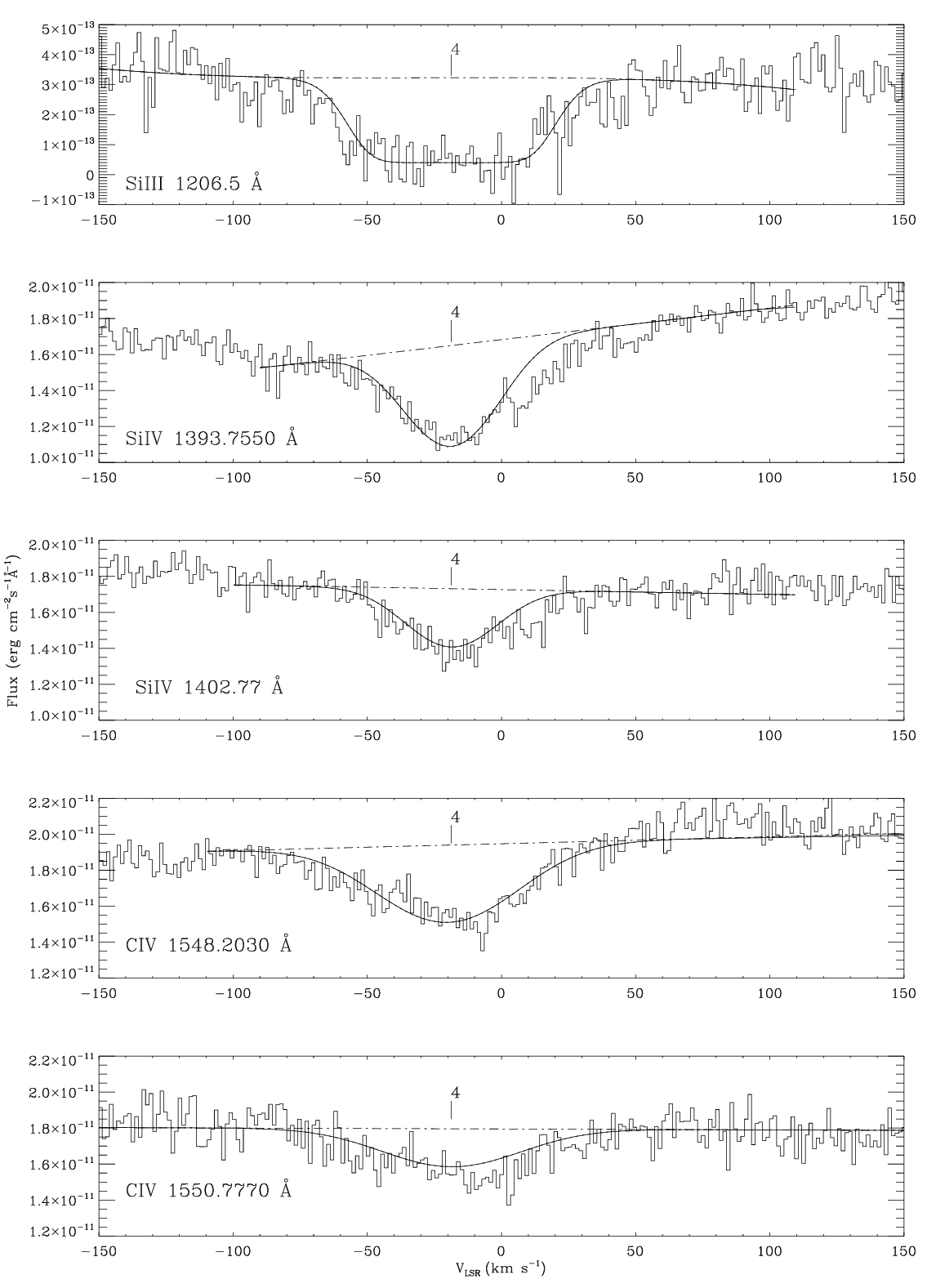}
 \end{center}
\caption{Highly-ionized species observed with STIS.  
As an indication, consistent absorption line profiles for gas at a 
temperature of 900\,000K and at the velocity of  component 4 are superimposed.
 \label{fig:highions}}
\end{figure}

{        We identify four groups of species,  classified according to the characteristics of their line
 profiles, which reflects  the distribution and the nature of the
gas traced by the given species. Examples of the line profiles are
presented in Figures~\ref{fig:molecules} to~\ref{fig:highions}. 
The figures also show the velocity components  and  fits to the profiles, which are the results of 
the analysis which will be
 described in Section~\ref{sec:data-analysis} after presentation of the complementary data in
Section~\ref{sec:complementary-data}. }

(1)~ {\large Molecules:} see Figure~\ref{fig:molecules}.
Several CO electronic A-X bands are present in the HST/STIS spectra. We use the four following 
bands where CO is detected in several  lines from rotational levels 
J"=0, 1 and 2 in the HST/STIS observations : A-X (0,0) at $1544.54\,\textrm{\AA}$, A-X (1,0) at $1509.9\,\textrm{\AA}$,
A-X (2,0) at $1477.57\,\textrm{\AA}$ and A-X (3,0) at 
$1447.4\,\textrm{\AA}$.
 {        Several other A-X bands  are present in the spectra but they are both weaker and noisier 
 (due to their location at shorter wavelengths hence in a fainter part of the spectrum). }
 The bands are resolved  into 
R and Q lines from the three rotational levels $J"=0$, 1 and 2, as illustrated in 
 Figure~\ref{fig:molecules} where two bands are shown as examples.

(2)~ {\large Neutral species with  ionization potential lower than that of H~{\sc i}:} see Figure~\ref{fig:neutrals}.\\
Atomic carbon lines have been measured in the three fine-structure
levels of the $^{3}P$ fundamental level: $J=0$ (hereafter {\rm {C~{\sc i}}}),
$J=1$ (hereafter {\rm {C~{\sc i}}*}), and $J=2$ (hereafter {\rm {C~{\sc i}}**}).
36 carbon lines have been detected, { among which we have selected 20 for use in the fits, 
eliminating  lines which  are either heavily blended with another line, or lack reliable f-values.} A 
representative subset of those lines is shown in Figure~\ref{fig:neutrals}, with
 lines from the other neutral species, Fe and S. These neutral lines are
narrow with a width of only a few resolution elements ({\it i.e.} a few
$\kms $), and clearly asymmetric indicating the presence of two velocity components
referred to as A and B.

(3)~ {\large Low ionization species and neutrals with ionization potential higher than  that of H~{\sc i}:} see 
Figure~\ref{fig:ions}.  \\
In those species the absorption extends to negative 
velocities as low as $ -50 \kms$ for some of the lines. 
The broad absorption is 
resolved 
into four distinct components. They are particularly well-resolved in the lines of the heavy element Fe~{\sc ii}.
These components are numbered 1 to 4 in order of decreasing velocity.

(4)~ {\large High ionization species:} see Figure~\ref{fig:highions}.\\
High ionization species Si~{\sc iii}, Si~{\sc iv}, C~{\sc iv} are detected in this line of sight. 
Their profile is broad, and centered at negative velocity close to  
the most negative velocity component seen in the other ions, Component 4.

\section{Complementary observations}\label{sec:complementary-data}
We present additional data sets  that  complement our STIS observations.
\subsection{Optical absorption lines \label{sub:Optical-Spectroscopy}}
Optical spectra of the ${\rm {CH}}$ and ${\rm {CH}^{+}}$ lines
about $4300\, \textrm{\AA}$  (see Table~\ref{Tab:Optical-absorption-lines}), 
the C$_2$ $\rm A ^1\Pi_u-X^1\Sigma_g^+(2,0)$ 
and CN $\rm A ^2\Pi-X^2\Sigma^+(2,0)$  molecular bands about
$8760$ and $7905 \, \AA$,
were obtained in April 1990 using the Coude
Echelle Spectrometer at the $3.6 \,$m of ESO.\\
\begin{table}[!htbp]
\begin{center}{\footnotesize }\begin{tabular}{c|c|c}
\hline
{\footnotesize Spec.}&{\footnotesize ${\rm CH}$}&{\footnotesize ${\rm CH^{+}}$}\tabularnewline
\hline
\hline
{\footnotesize Band}&{\footnotesize ${\rm A}^{2}\Delta\!\!-\!\!{\rm X}^{2}\Pi\,(0,0)$}&
{\footnotesize ${\rm A}^{1}\Pi\!-\!{\rm X}^{1}\Sigma^{+}\,(0,0)$}\tabularnewline
\hline
{\footnotesize Line}&{\footnotesize ${\rm R_{2e}(1)\!+\! R_{2f}(1)}$}&{\footnotesize ${\rm R(0)}$}
\tabularnewline
\hline
{\footnotesize $\lambda(\textrm{\AA})$}&
{\footnotesize $4300.313$}&
{\footnotesize $4232.548$}
\tabularnewline
\hline
{\footnotesize $f$}& 
{\footnotesize $5.06\,10^{-3}$}&
{\footnotesize $5.45\,10^{-3}$}
\tabularnewline
\hline
\end{tabular}\end{center}{\footnotesize \par}
\caption{Molecular absorption lines detected in the optical spectra. Listed parameters are from Gredel et al. (1993).
\label{Tab:Optical-absorption-lines}}
\end{table}

 The spectrograph was
connected to the $3.6 \,$m beam by a fiber link. The light coming from the
fiber was divided in 10 independent slices. Each of the slices
produced a 20-pixel wide spectrum on the CCD. To reduce the read-out
noise CCD pixels were read in bins of 10 pixels, perpendicular to the
dispersion direction. The optics of the spectrometer were oriented so that
the spectral resolution was not reduced (i.e. the dispersion
direction was perpendicular to the binning direction). The
resolving power was 110 000 corresponding to a velocity resolution of
$2.7 \kms $. The sampling was 2.2 pixels per resolution
element, and the spectra were 1024 pixels in length. Wavelength calibration
was performed using an observation of a Th-Ar cathode
lamp. This calibration was repeated for each set-up of the spectrometer.
The identification and line-center measurements of a set of around 20 lines in these
spectra allowed us to derive the dispersion relation with an accuracy
better than 10~${\rm {m}}\textrm{\AA}$.  Flat-field
images were acquired at the beginning and end of each set of observations,
for the entire wavelength range. 

The images were corrected for dark current, and then flat-field corrected. Spectra were extracted from the images 
by averaging  20 lines
over which the flux is spread. The spectra were normalized
by dividing the original spectra by a median average showing no spectral lines from 
interstellar gas or the Earth atmosphere.  Atmospheric lines were identified using  observations of 
HD~106911, a bright star in Chamaeleon with almost no reddening.   \\
Sections of the normalized spectra that include  the  CH and CH$^+$ lines, 
are presented in Figure~\ref{fig:molecules}. 
{        The C$_2$ and CN spectra show no detections with $3 \sigma $ upper limits 
of 2.7 and $1.0 \,$m$\AA$, respectively. }
Gaussian fits to the  CH and CH$^{+}$ lines have been used to derive 
the line parameters, and thereby the
column density $N$. Because the transitions are optically-thin,  $N$
is related to the measured
equivalent width, $W_{\lambda}$, by the equation: $ N=1.13\,10^{20}\,W_{\lambda}/{\lambda^{2}\, f}$
with $W_{\lambda}$ and $\lambda$ in $\textrm{\AA}$, $N$ in ${\rm
cm^{-2}}$ and where $f$ is the line oscillator strength given in 
Table~\ref{Tab:Optical-absorption-lines}. 
Results are presented in Table~\ref{tab:abscolumn}.
{        The optical absorption lines are also fitted, together with the UV 
absorption lines, in the general spectral analysis described in Section~\ref{sub:molecules}.
This, in particular, allows  the CH and CH$^+$ velocity distribution to be related to the velocity 
structure seen in other species.   }
\subsection{H~{\sc i} emission \label{sub:HI-emission-spectra}}
A sensitive 21-cm H~{\sc i} southern sky survey has recently been published (Bajaja et al. 2005). 
Its angular resolution is half a degree and the velocity resolution $1.3 \kms $. 
\begin{figure}[!htbp]
\begin{center}
\includegraphics[width=0.95\columnwidth]{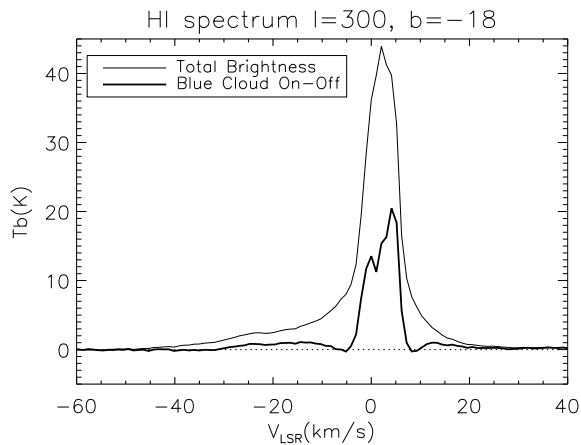}
\end{center}
\caption{H~{\sc i} emission line in the direction of HD~102065 derived 
from the IAR Southern Sky survey. The thick line represents the difference with respect to
a reference position outside of the Blue cloud.
\label{HI_spectrum}}
\end{figure}
In Figure~\ref{HI_spectrum}, we present the survey spectrum for the position closest to
HD~102065 (thin line). 
The thick line is the difference between this spectrum and  an 
OFF cloud spectrum obtained by averaging 3 spectra  outside the Blue Cloud (Dcld 300.2-16.9)
at $\rm l=297^\circ$ to $297.5^\circ$ and $\rm b=-19.5^\circ$ to $-19^\circ$, corresponding to a 
local minimum in the $100\mu$m IRAS map. 
This subtraction  is supposed to remove
the background emission, but may also remove foreground emission that is not
spatially correlated with the Blue Cloud.
Within 
unknown variations of the background emission 
between the ON and OFF cloud positions, 
this difference spectrum represents the H~{\sc i} spectrum of the Blue Cloud. 
In Table~\ref{Tab_emission}, we list the results of a Gaussian decomposition of this spectrum.
The column densities have been calculated with  
the assumption that the H~{\sc i} emission line is optically-thin.

\begin{figure}[!htbp]
\begin{center}
\includegraphics[width=0.95\columnwidth]{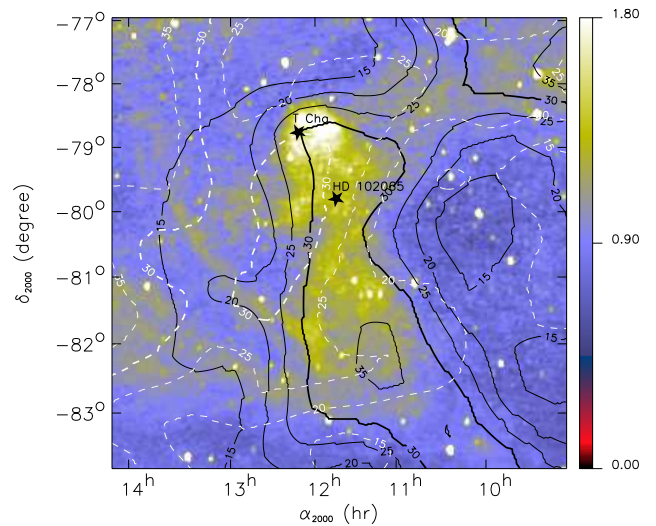}
\end{center}
\caption{IRAS $12\mu m$ image (IRIS processing,  Miville-Desch\^{e}nes and Lagache, 2005)
with contours of H~{\sc i} emission at $ \rm V_{LSR} = -1 \kms $ (white dashed) and 
$5 \kms $ (black solid) taken from  the IAR survey
(30' resolution and  $1.3 \kms $ width velocity channels, Bajaja et al. 2005). 
The position of HD 102065 and the T Tauri star T Cha
are marked. Contours are labeled with the brightness temperature. The 30~K contour is thicker.
The cloud tail is roughly at a constant Galactic longitude. The Galactic coordinates of
HD~102065 are $l\, =\, 300.0^\circ$ and $b\, =\, -18.0^\circ$. The electronic version is in color.
\label{Cham12_HI} }
\end{figure}
\begin{figure}[!htbp]
\begin{center}
\includegraphics[width=0.95\columnwidth]{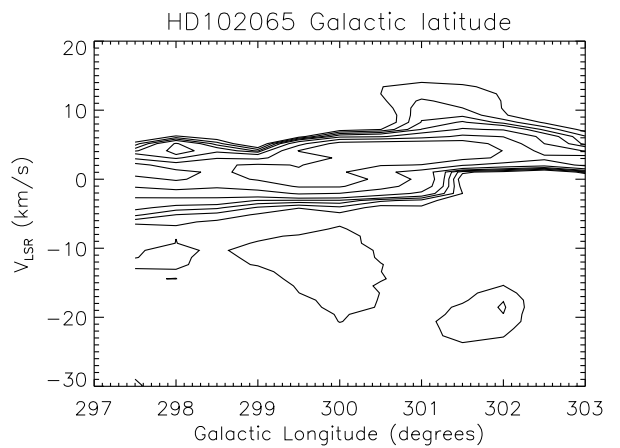}
\end{center}
\caption{H~{\sc i} position-velocity diagram across the Blue Cloud 
at the Galactic latitude of HD~102065 (b=-18$^\circ$). This figure shows that the two velocity
components at respectively 0.1 and $4.2 \kms $ each extends on one side of the star position (l=300$^\circ $) 
where they overlap. 
The emission at the OFF position (see text) has been subtracted to highlight the
gas associated with  the cloud.
\label{Fig:HD_HI_LV} }
\end{figure}

In Figure~\ref{Cham12_HI}, we show the spatial distribution of the H~{\sc i} emission in two-channel maps 
($\rm V_{LSR}\, =\, -1$ and $+5 \kms $)
overlaid over the IRAS 12$\mu$m image (Miville-Desch\^{e}nes and Lagache, 2005), and in 
Figure~\ref{Fig:HD_HI_LV} a position-velocity diagram across the Blue cloud 
at the Galactic latitude of HD~102065.  
These two figures show
that the Blue Cloud, the bright IRAS 12$\, \mu$m emission on the map, is 
associated with gas at low velocities. 
The total brightness spectrum (thin line in Figure~\ref{HI_spectrum}) shows the
presence of a broad, negative velocity component, extending down to $-40 \kms $.  
This gas is not spatially 
correlated with the infrared emission from the Blue Cloud. 
It is however observed  in UV-absorption spectra, and must therefore be located in front of the star.
\begin{table*}[!htbp]
\begin{center}\begin{tabular}{l|ccccc}
\hline
species&
$ \rm V_{LSR}\,\,  ({\rm {km\, s}^{-1}}) $&
$\rm T_b\,\,{\rm (K)}$&
$ \rm FWHM\,\,({\rm {km\, s}^{-1}})$&
$\rm b\,\,({\rm {km\, s}^{-1}})$&
$\rm N\,\,({\rm {cm}^{-2}})$\tabularnewline
\hline
H~{\sc i} \, 21 cm (difference)&$14.5\pm0.6$&$0.9\pm0.15$&$7.3\pm1.4$&$4.4\pm0.8$&$1.3\pm0.2\,10^{19}$\tabularnewline
&$4.2\pm0.1$   &$20.4\pm1.1$ &$3.2\pm0.2$ &$1.9\pm0.1$ &$1.3\pm0.07\,10^{20}$\tabularnewline
&$0.1\pm0.1$   &$13.4\pm0.7$ &$4.3\pm0.3$ &$2.6\pm0.2$ &$1.1\pm0.06\,10^{20}$\tabularnewline
&$-17.5\pm0.6$ &$1.0\pm0.07$ &$16\pm1.4$  &$9.4\pm0.8$ &$3.0\pm0.2\,10^{19}$\tabularnewline

\hline
CO(1-0) &$2.8\pm0.2$&$7.0\,10^{-2}$&$3.5\pm0.3$&$2.1\pm 0.2$& $6.9\,10^{13} ~ (J\, = \, 1)$\tabularnewline
CO(2-1) &$2.75\pm0.17$&$3.6\,10^{-2}$&$3.0\pm0.3$&$1.8\pm0.2$&$1.8\,10^{13} ~ (J\, =\, 2)$\tabularnewline
\hline
\hline
 & &$I\,({\rm erg\, s}^{-1}{\rm cm}^{-2}{\rm sr}^{-1})$ & &
&N({\rm {C~{\sc II}}*})\tabularnewline
\hline
${\rm C^{+}\, 158\mu m}$&-&$2.8\,10^{-6}$&-&-&$1.2\,10^{15}$\tabularnewline
\hline
\end{tabular}\end{center}
\caption{Results from emission line measurements.
H~{\sc i}  measurements are performed on the difference spectrum 
(see text and Figure~\ref{HI_spectrum}).
The H~{\sc i}
component at $14.5 \kms $ is not detected in absorption and is
probably due to background gas situated behind the star HD~102065.
 \label{Tab_emission}}
\end{table*}
\subsection{CO emission spectra\label{sub:CO-emission-spectra}}
Long integrations have been performed with the 15m SEST telescope on the 
$^{12}$CO $J$=2-1 and  $J$=1-0 emission lines.\\
The $J=$1-0 and 2-1 lines were observed in 1990 and 2001, 
with system temperatures
of 700 K and 200 K respectively, and a velocity resolution of $0.1\kms $.   
The rms noise level of the $^{12}$CO $J=$2-1 line displayed in 
Figure~\ref{fig:molecules}
is sufficiently low  ($\sim 10$ mK)  to reveal a broad and weak
component ($ \Delta v \, = \, 2.9\kms $, $T_A ^* \, =\, 36\, mK$), and a
marginally-detected narrower component 
($\Delta v \, = \, 0.5\kms $, $T_A ^* \, = \, 15 \, mK$).
The antenna temperature has not been corrected for beam coupling to
 the sky, because the source is much more extended than 
the telescope-beam size at
 this frequency ($\sim 22"$ or 0.015 pc at the distance of the
 cloud). This is not a critical issue because the beam efficiency
 was $\sim 0.85$ at 230 GHz. 
Similarly, the rms noise level of the $J=$1-0 line is $\rm 18\, mK$ for a
velocity resolution of $0.2 \kms $.
Given the weakness of the $^{12}$CO  lines, the assumption of
optical thinness is justified and the column densities of
molecules in the $J=2$ and 1 levels are inferred from the line integrated
intensities $ W(CO)\, =\, 0.1$ and $0.008 \, K \kms $  of the broad and
narrow components of the $J=2-1$ line, and $W(CO)\, = \, 0.2\, K \kms $ for the J=1-0 line.
The respective column densities are given in Table~\ref{Tab_emission}.
\subsection{C$^+$ observations\label{sub:C+-obs}}
The ${\rm {C^{+}}}$ fine-structure transition at ${\rm {158\, \mu m}}$ has
been measured towards HD~102065 with the Long Wavelength Spectrometer
(LWS) of the Infrared Space Observatory (ISO), with a spectral
resolution of $\rm R\, = \, 300$, much lower than
all other data presented in this paper.  The data have been processed
following the LWS handbook (Gry et al 2003).  The resulting line
intensity  I(C$^{+}$ $\lambda158\mu $m) is $\rm 2.8\pm 1.0\, 10^{-6}\,{erg\,
s^{-1}\, cm^{-2}\, sr^{-1}}$, from which we estimate the
C~{\sc ii}* column density  N(C~{\sc ii}*)$\, = \, 1.2 \pm 0.4\,10^{15}\,
cm^{-2}$.  This is a measurement of the mean column density over the
80" LWS beam.

\section{Analysis of the line of sight - Description of the interstellar components\label{sec:data-analysis}}
\subsection{The tool: Multi-lines fitting\label{sub:line-fitting}}
To derive the characteristics of the interstellar components,
i.e. column density $N $, velocity $ v $ and broadening parameter $b $, 
we fit the observed line-profiles with theoretical ones, which are the results
of the convolution of a Voigt profile with the instrumental function
(LSF). We use a profile-fitting method, instead of a traditional measurement of equivalent width, because 
the velocity structure is complex as can be seen in Figures ~\ref{fig:molecules} to \ref{fig:ions}.
Because several components are blended
together in most lines, profile fitting is required to derive the
characteristics of the individual components.  The data fitting has
been performed with the use of the profile fitting software
{}``Owens.f'' developed by Martin Lemoine.

The software allows  several lines of the same element,
as well as lines from different elements, to be fitted together. It also allows 
several velocity components to be fitted simultaneously, and the
components to be included in the absorption profiles for each
element to be specified.
The software therefore derives a global and consistent
solution for a group of species.  Multiple iterations are necessary to
find which species could be fitted in a consistent way with the same
velocity components. A common component for different species
obviously implies a common absorption velocity but also common
physical conditions, implying consistent broadening parameters.

The fitting software breaks the line-broadening parameter (b-value)
into thermal broadening, which depends on the element mass, and
non-thermal broadening (turbulence), which is the same for all elements
in a component.  
Therefore, fitting lines from elements of different masses simultaneously, in principle enables the 
simultaneous measurement of temperature, turbulent velocity, the velocity itself, and column densities.
 In practice interstellar absorption lines are usually found to be of two types.
 The first type displays  unsaturated lines from heavy elements that have similar b-values 
 independently of the element mass: in this case, the broadening is primarily non-thermal, implying that T is low and  
 can be estimated only if the analysis includes an element whose mass is low enough to introduce a significantly different 
 b-value. Here we will consider H~{\sc i} emission lines. In the case of the second type of interstellar absorption line, 
 the b-values vary with element mass due to a significant thermal effect, implying that T is high and can be estimated.
In practice the interstellar absorption lines  are broader than the 
resolution element (for T  higher than a few thousand K), and strong lines are occasionally 
broadened considerably by saturation, leading to the blending of components  and limiting the  precision with which broadening 
parameters can be derived.

\begin{figure}[!htbp]
\begin{center}\includegraphics[width=0.6\columnwidth,angle=-90]{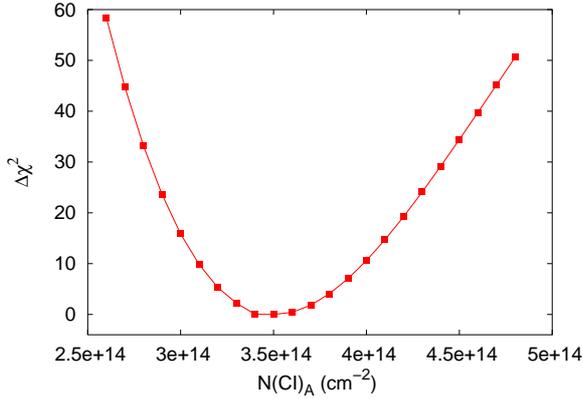}\end{center}
\caption{Example of a $\Delta\chi^{2}$ curve, here for C~{\sc i} in component A. Each point
corresponds to a fit performed with the C~{\sc i} column density in component A fixed to the value 
given in abscissa.  $\Delta\chi^{2}$ is rescaled by  dividing it by the 95 \%-confidence level for this fit (rescaled
$\Delta\chi^{2}=1.41$). 3$\sigma$-uncertainty is given by $\Delta\chi^{2}=3^{2}\times1.41=12.7$.\label{chi}}
\end{figure}

The quality of the fits, and  the 
error bars on column densities, have been computed using the
$\Delta\chi^{2}$ method described, for example, in H\'ebrard et al (2002). We performed
several fits by fixing the column density of a given element ${\rm X}$
to a different value in each fit. In all of these fits, all other
parameters are not fixed and we compute the best $\chi^{2}$ for each value of
column density. The plot of $\Delta\chi^{2}$ versus $N(X)$ provides the
1, 2, 3, 4, $5 \, \sigma$ range for $N({\rm X})$. This is illustrated
with an example in Figure~\ref{chi}.
 Examples of the fits
are presented in Figure~\ref{fig:molecules} for molecules, Figure~\ref{fig:neutrals} for neutrals,
and Figure~\ref{fig:ions} for ions.
\subsection{Component structure and velocities\label{sub:structure}}
\subsubsection{Atoms with ionization potential lower than H~{\sc i}.}
The common asymmetry of the absorption lines of the neutral species
(Figure~\ref{fig:neutrals}) suggests the existence of two velocity
components in all of these lines.  When fitting simultaneously 
C~{\sc i}, C~{\sc i}*, C~{\sc i}**, S~{\sc i} and 
Fe~{\sc i}, a best fit is achieved for two components, that we
call component A and component B, for the following LSR velocities:
$\rm V_{LSR}=3.8$ and $0.4 \kms $. They may be converted into heliocentric
velocities by adding $10.2\kms $.

The b-values derived in the fits, $b_{{\rm A}}=1.5\pm0.2$ 
and $b_{{\rm B}}=1.6\pm0.2 \kms $
are not significantly different for the three species with different
masses. This is  strong support to the fact that broadening is
primarily non-thermal. 

Interestingly,  the velocities identified for the two main components  in the H~{\sc i} Blue Cloud
emission spectrum (see Table~\ref{Tab_emission}) are identical to those of A and B within the error bars. 
The b-values are also similar to those expected in cool gas, with the H~{\sc i} b-values being slightly higher  than that 
of S~{\sc i} and C~{\sc i} due to the mass difference.
 The main components seen in emission
are therefore likely to be related to the two main components observed in absorption.
\subsubsection{ Molecules}\label{sub:molecules}
In contrast, the CO lines do not accept a two-component solution. 
{        The singularity of the CO component is evidenced  by the  small width of 
the CO lines, with $b_{CO} =1.6\pm0.2\,{\rm km\, s^{-1}}$,
similar to that of component A or B alone. This width is  significantly lower than the 
velocity difference between A and B, $3.3\,{\rm {km\, s}}^{-1}$, 
toward which  value $b_{CO}$ would tend to increase if CO was distributed over the two components. 
When fitting the ${\rm {CO}}$ lines in a two-component model together with all well-behaved 
(i.e. not heavily saturated) neutral lines, unavoidably in the resulting fit most of the ${\rm {CO}}$ column density is found
in one of the components, and the resulting theoretical profile appears slightly shifted
relative to the observations. 
This component cannot be identified with either A or B, since its velocity shift is 
significantly higher than the uncertainty of the wavelength calibration in a single STIS sub-spectrum. 
Alternatively, when fitting the neutrals
and ${\rm {CO}}$ lines altogether in a one-component model, the fit
to the neutral lines is significantly degraded. 
The line characteristics (central velocity and line width) 
of CO in emission agree  
with that derived from UV absorption lines. 
This agreement validates
the UV wavelength calibration and  confirms that CO,  
unlike H~{\sc i} and the atomic species C~{\sc i} and S~{\sc i}, is found at 
a single velocity.

We infer that  ${\rm {CO}}$ is detected primarily in one  component at a velocity intermediate to that 
of A and B,  $\rm V_{LSR}{\rm CO}=2.7\,{\rm {km\, s}^{-1}}$. We will call this component C
(see Figure~\ref{fig:molecules} illustrating  the line fit  for two CO bands).}

Further information is provided by the CH and CH$^+$ absorption line
profiles.  We performed the profile fitting of the CH and CH$^+$
lines with Owens together with the STIS data of the species C~{\sc i}, S~{\sc i} and
CO.  We  made several fits with CH and CH$^+$ either using one or two
components, and with a velocity shift as a free parameter to take into account
uncertainties in the absolute wavelength calibration of both the UV and optical spectra. 
The best results have been obtained if {\it (i)} CH$^+$ is present in both
components A and B and {\it (ii)} most of the CH absorption is in one
component, corresponding to the CO component C.  This solution implies
that the optical spectra are shifted by about 0.3  $\kms $ relative to
the STIS data, a value consistent with the wavelength calibration uncertainty.

We conclude that the bulk of the CO and CH molecules are in a single component (C) at a different velocity 
to the bulk of the neutral atoms carbon, sulfur, iron and hydrogen and the CH$^{+}$ molecules, which appear
to be distributed over
two well-separated velocity components (A and B). {        Of course it is not excluded  that some 
neutrals are present at the velocity of CO, and alternatively that some CO is present in the components A and B, 
at  undetectable levels. We believe that the three components A, B and C are, in fact, physically 
related. Within this assumption, the three velocity components would reflect the particular velocity structure 
associated with local abundance variations within a  {\it single cloud}. }

\subsubsection{ Ions and neutrals with ionization potential higher than H~{\sc i}.}
Almost all ions and neutrals with an ionization potential slightly higher
than H~{\sc i}: Mg~{\sc ii}, Mg~{\sc i}, Mn~{\sc ii}, 
P~{\sc ii}, Ni~{\sc ii}, C~{\sc ii}, N~{\sc i} and O~{\sc i},
can be fitted altogether with a unique solution for the line of sight
structure : component 1 at $2.5\pm0.3\,{\rm {km\,
s}^{-1}}$, component 2 at $-5.5\pm0.2\,{\rm {km\, s}^{-1}}$, component
3 at $-11.0\pm0.2\,{\rm {km\, s}^{-1}}$ and component 4 at
$-18.7\pm0.2\,{\rm {km\, s}^{-1}}$ (velocities are LSR, see Table~\ref{tab:abscolumn}).

The low-velocity component, component 1, corresponds to the gas 
detected in  neutral and molecular lines. In the
unsaturated ion lines  this component can be fitted either with two
components at the velocity of A and B, or with
a single component at the intermediate velocity $\rm V_{LSR}\,=\,2.5\,\pm\,0.3\,{\rm {km\,
s}^{-1}}$ 
without affecting the column density results or the quality of the
fits (monitored by the $\chi^{2}$). 
However when adding stronger lines to the fit, it has been impossible
to find a good fit to the spectra in component 1 with the column density
provided by the unsaturated lines and the A+B velocity structure. This
suggests that the structure of this component is more complex and  may indicate the presence 
of warmer gas at the velocity of component 1.
Even with a low column density, in strong lines  warm gas can contribute  a large part of
the equivalent width due to a  larger b-value.
For the ions, we therefore
quote a unique column density, obtained using the unsaturated lines
when they exist, for the low velocity component 1 alone.\\
The components structure is summarized in Table~\ref{tab:abscolumn}. 
\subsection{Column densities\label{sub:columndensity}}
\subsubsection{Column densities from absorption lines}
Once the velocity structure is defined, column densities
can be derived  following the method described in section~\ref{sub:line-fitting}. 
In this respect, not all of the lines are equally useful.

Fe~{\sc ii} is the most favorable 
ion in our sample for two reasons. First,
several Fe~{\sc ii} lines are available with a large range of
f-values  from 0.0014 to 0.32 (see Table~\ref{fvalues}). Strong lines allow detection of weak 
components, and weak lines enable  non-saturated profiles of  strong components to be analyzed. Another
advantage of Fe~{\sc ii} lines is that, due to its high mass, Fe is less affected by
thermal broadening than lighter elements;
therefore in warm gas Fe~{\sc ii} absorption lines are narrower and the
components better-resolved as seen on Figure~\ref{fig:ions}. The errors in N(Fe~{\sc ii}) are therefore due only  to
the combination of noise and uncertainty in the continuum placement.

We achieve our poorest quality analyses for ions for which only strong
lines are available. This is the case here for C~{\sc ii}, C~{\sc ii}$^*$ and Si~{\sc ii}, for
which  all components are saturated and for which we can derive only
lower limits.  

When only one weak line  for a given
element is detected, only the strongest component 1 is detected and measured and we
get column-density upper limits for all other components. This is the case for both Ni~{\sc ii} and P~{\sc ii}.

For some elements, we have both a very weak line, which is only detected in
the strongest component, allowing  measurement of the column density, and very strong
lines that are saturated in all components. Combining weak and strong lines gives upper and lower
limits for the weak components. This is the case for 
Mg~{\sc ii} $\lambda1240\textrm{\AA}$ and  O~{\sc i} $\lambda1356\textrm{\AA}$, which
are respectively $4.4\times10^{4}$ and $10^{3}$ times weaker than the
strong lines  Mg~{\sc ii} $\lambda2800\textrm{\AA}$ and
O~{\sc i} $\lambda1302\textrm{\AA}$.  S~{\sc ii} is an intermediate case where the
three S~{\sc ii} lines are saturated in  component 1  and at least one S~{\sc ii} line is unsaturated in the other
weaker components.  The
uncertainty in their column density is due to blending between
components and uncertainty in the continuum placement. 

 As mentioned in Section 2.2,  Si~{\sc ii}* and O~{\sc i}** are detected primarily  at the 
 velocity of component 4, while they are hardly detectable in
components~1, 2 and 3.  Nevertheless O~{\sc i}*
is detected in the four components.

Column densities resulting from absorption line fitting  are provided in Table~\ref{tab:abscolumn}.
\begin{table*}[!htbp]
\begin{center}\begin{tabular}{lcccccc}
\hline
Component&A&C&B&&&
\tabularnewline
$\rm V_{ LSR}\,({\rm {km\, s^{-1}}})$&3.8$\pm0.1$&2.7$\pm0.2$&0.4$\pm0.1$&&&\tabularnewline
T (K) & $85\pm65$ & - &$270\pm100 $&&&\tabularnewline
$\rm b_{NT}~ (\kms )$ & 1.5$\pm0.2$&1.6$\pm 0.2$&1.5$\pm0.2$ &&&\tabularnewline
\hline
\\[-1ex]${\rm {CO}}\, J=0$&
&{$3.3\pm_{0.5}^{0.3}\,(13)$}&&&&
\tabularnewline ${\rm {CO}}\, J=1$&
&{$2.1\pm_{0.3}^{0.4}\,(13)$}&&&&
\tabularnewline ${\rm {CO}}\, J=2$&
&{$<2.5\,(12)$}&&&&
\tabularnewline ${\rm CH}$ & &$6.5\pm1.5\,(12) $&& &&
\tabularnewline ${\rm CH^+}$ & $9\pm2\,(12)$&&$4\pm1\,(12)$& &&
\tabularnewline C~{\sc i} ($^{3}P_{0}$)&$3.5\pm_{0.3}^{0.5}\,(14)$&&$1.3\pm_{0.3}^{0.2}\,(14)$&&&
\tabularnewline C~{\sc i}* ($^{3}P_{1}$)&$6.6\pm_{0.7}^{0.8}\,(13)$&&$2.7\pm0.5\,(13)$&&&
\tabularnewline C~{\sc i}**($^{3}P_{2}$)&$1.1\pm0.15\,(13)$&&$3.0\pm_{1.1}^{2.2}\,(12)$ &&&
\tabularnewline S~{\sc i}&$4.4\pm_{0.5}^{0.6}\,(12)$&&$2.0\pm0.5\,(12)$&&&
\tabularnewline
Fe~{\sc i}& $1.50\pm1.0\,(11)$&& $<2.81\,(10)$&&&\tabularnewline
\\
\hline
\hline
Component&
\multicolumn{3}{c}{1 (A+B)}& 2& 3& 4\tabularnewline
$\rm V_{{\rm LSR}}\,({\rm {km\, s^{-1}}})$& \multicolumn{3}{c}{$2.5\pm0.3$}& $-5.5\pm0.2$ 
& $-11.0\pm0.2$& $-18.7\pm0.2$\tabularnewline
T (K)&\multicolumn{3}{c}{-}&8\,500 $\pm 1\,500$& 13\,000$\pm 3\,000$& $\sim 20\,000 - \geq 200\,000$\tabularnewline
\hline
\\[-1ex]Fe~{\sc ii}&\multicolumn{3}{c}{$2.2\pm\,1.1\,(14)$}&
$1.2\pm_{0.3}^{0.1}\,(13)$&$1.3\pm0.2\,(13)$&$1.4\pm0.2\,(13)$\tabularnewline
Mg~{\sc ii}&\multicolumn{3}{c}{$1.8\pm_{0.5}^{0.3}\,(15)$}&
$1.9\,(12)\mapsto1.2\,(14)$ &$2.3\,(12)\mapsto1.5\,(14)$ &$2.9\,(12)\mapsto1.6\,(14)$\tabularnewline
S~{\sc ii}&
\multicolumn{3}{c}{$>3.0\,(16)$}&$5.0\pm_{1.5}^{2.0}\,(13)$&$1.0\pm0.2\,(13)$&$1.1\pm_{0.2}^{0.4}\,(14)$\tabularnewline
P~{\sc ii}&
\multicolumn{3}{c}{$2.5\pm1.0\,(14)$}&$<1.6\,(13)$&$<2.4\,(13)$&$<2.7\,(13)$\tabularnewline
Mn~{\sc ii}&
\multicolumn{3}{c}{$1.3\pm_{0.1}^{0.2}\,(13)$}&$1.2\pm1.0\,(11)$ &$1.6\pm1.0\,(11)$&$1.4\pm1.0\,(11)$\tabularnewline
 C~{\sc ii}&
\multicolumn{5}{c}{$>1.3\,(14)$}\tabularnewline
 C~{\sc ii}*&
\multicolumn{5}{c}{$>1.3\,(14)$}\tabularnewline
 Si~{\sc ii}&\multicolumn{5}{c}{$>1.0\,(14)$}\tabularnewline
Si~{\sc ii}*&\multicolumn{3}{c}{$<2.4\,(12)$}&$<1.6\,(12)$&$<1.8\,(12)$&$>3.2\,(13)$\tabularnewline
Ni~{\sc ii}&\multicolumn{3}{c}{$7.0\pm2\,(12)$}&$<1.6\,(13)$&$<1.2\,(13)$&$<1.4\,(13)$\tabularnewline
Mg~{\sc i}&\multicolumn{3}{c}{$2.25\pm_{0.28}^{0.5}\,(12)$}&$<2.1\,(11)$&$<2.1\,(11)$&$7\pm4$$\,(10)$\tabularnewline
O~{\sc i}&\multicolumn{3}{c}{$4.1\pm 1.6\,(17)$}&$3.2\,(13)\mapsto6.2\,(16)$&$3.4\,(13)\mapsto4\,(16)$&$3.1\,(13)\mapsto3.7\,(16)$\tabularnewline
N~{\sc i}&\multicolumn{3}{c}{$2.0\pm1.0\,(17)$}&\textrm{$1.5\,(14)\mapsto1.2\,(16)$}&\textrm{$2.7\,(14)\mapsto1.9\,(16)$}&\textrm{$3.0\,(14)\mapsto1.2\,(17)$}\tabularnewline
S~{\sc iii}&\multicolumn{3}{c}{}&&&$<2.4\,(13)$\tabularnewline
Si~{\sc iii}&\multicolumn{3}{c}{}&&&$>6\,(12)$\tabularnewline
Si~{\sc iv}&\multicolumn{3}{c}{}&&&$8.2\pm3.2\,(12)$\tabularnewline
C~{\sc iv}&\multicolumn{3}{c}{}&&&$1.45\pm0.6\,(13)$\tabularnewline
\hline
\end{tabular}\end{center}
\caption{\label{tab:abscolumn}Interstellar component characteristics 
derived from the absorption lines in the UV and optical
spectrum. The ions absorption in  component 1 may be distributed 
over the velocities of components A and B however the fits have been performed with one component. 
{\bf Velocities:} The listed uncertainties on $V_{LSR}$ apply to relative
velocities. The absolute uncertainty from the precision of the STIS
wavelength calibration is about $1.2\,{\rm {km\, s}}^{-1}$,  absolute velocities have however been 
confirmed to about $0.3\,{\rm {km\, s}}^{-1}$ by the good agreement between component
C and the CO emission in the heterodyne data, which have an absolute
precision of $0.1\,{\rm {km\, s}}^{-1}$.  
{\bf Temperature:} When possible, temperatures are derived from  the combination of
$b$-values from species of different masses. The precise temperatures for components A and B 
have been derived under the assumption that they can be identified to the two components seen in 
emission in the Blue Cloud H~{\sc i} spectrum.  
{\bf Column densities} (in cm$^{-2}$): Numbers in parentheses are powers of 10. All error bars are
3 $\sigma$ uncertainties, except for the intervals given by the
arrows, indicating and upper limit derived from a faint line and a
lower limit derived from a strong saturated line. Si~{\sc ii},
C~{\sc ii} and C~{\sc ii}* profiles are saturated for all
components, implying that only a lower limit could be derived for the whole
absorption.}
\end{table*}
\subsubsection{Comparison of absorption/emission column densities}

In this section, we compare column densities derived 
from emission and absorption spectra.  

The total hydrogen column density can be estimated from the color excess,  and the standard ratio 
$N({\rm H}_{tot})/E({\rm {B-V}})=5.8\times 10^{21}\,{\rm {cm}}^{-2}\,{\rm{mag}}^{-1}$ \citep{BSD78}.
This provides a measurement of $N({\rm H}_{tot})=9.8 \times 10^{20}$ cm$^{-2}$. 
It can also be  derived in addition using H$_{2}$ column densities  measured in FUSE absorption spectra  
(Table~\ref{Tab_1}),  and the H~{\sc i} column density estimated from the Blue Cloud H~{\sc i} emission spectrum :
 $N({\rm H}_{tot})= N({\rm H})+2N({\rm H_2})= 9.6\times 10^{20} \, {\rm cm}^{-2}$ excluding the  
$14.5 \kms $ background component (Table~\ref{Tab_emission}). 
The quantitative agreement between emission and absorption results for  column densities, velocities, 
and line widths, supports the idea that the two H~{\sc i} low velocity
components observed in the Blue Cloud H~{\sc i} emission spectrum 
correspond to  components A and B observed in absorption.

One can also note  that there is a consistency between the C$^+$ column density derived
using ISO-LWSdata, and the lower limits measured from the saturated absorption lines.

CO column densities were measured both in emission and absorption.
 In absorption, they can be directly compared to the H$_{2}$ column densities derived from  
 FUSE spectra (Gry et al., 2002). This comparison enables the measurement of a total CO abundance
$N({\rm CO})/N({\rm H_2})\, = \, 1.5^{+0.2}_{-0.15}10^{-7}$.

It was interesting to find  that
 the CO column densities  were larger  when determined in emission than in absorption, by a
factor 3 in the $J\, =\, 1$, and by more than a factor of 7 for the $J\, =\,2$ level. 
Because the spatial resolutions of the observations are very different, this could imply
the existence of small-scale structure of the CO
emitting gas within the SEST beam. On the other hand, part of the CO emission 
could be background emission  from the envelope of the nearby Chamaeleon III cloud, 
which is observed over the same  velocity range \citep{Mizuno2001}.
Based on measurement of the CO excitation, we consider that this second possibility is however unlikely.  
The column density ratio between
the $J=2$ and 1 levels, derived from the emission spectrum 
provides an  excitation temperature $T_{ex}\, = \, 2.9\, $K
(population ratio $N(J=2)/N(J=1)=0.24$), 
a value very close to the Cosmic Microwave Background temperature. 
The low excitation implies that collision excitation of the CO transitions by H or
H$_2$, is negligible when  compared to radiative excitation, hence the local
density is much lower than the critical density of the 2-1 transition
$n_H=10^4\, $cm$^{-3}$. However, if the gas was associated with the 
Cham III cloud, we
expect that radiative excitation alone could produce a higher  $T_{ex}$.

\subsection{Temperatures}
As noted in Section~\ref{sub:line-fitting}, information on the 
temperature of the gas in each component follows from the ability to
fit simultaneously lines from several elements of different masses, which
allows the nature of the broadening to be determined.
%

\subsubsection{Cool gas}
For components A and B, 
the similarity of the $b$-values for lines from heavy elements 
of different atomic masses (see Section~\ref{sub:structure})
suggests that their widths are dominated by non-thermal motions.
To estimate the gas temperature, 
we  compared heavy-element line-widths to  the widths of the associated H~{\sc i} line 
emission. 

Although the beam of the H~{\sc i} emission observations is several orders of
magnitude larger than the pencil beam of  UV absorption
observations ($0.5^\circ$ vs milli-arcseconds), we assume that 
the samplings of the velocity
field are similar. By computing spectra produced by numerical
simulations of mildly compressible turbulence, Pety and Falgarone
(2000)  showed that  one-dimensional sampling of a turbulent,
homogeneous field is similar to that performed by an extended
 beam. This behavior of the linewidths 
is indeed found in mm-line observations 
of CO in emission and absorption against extragalactic sources 
\citep{Liszt98}.
Once the H~{\sc i} linewidths are included in the analysis,  
the temperatures, yielded formally by the $b$-value differences between
H~{\sc i} and heavy neutrals, are 85 $\pm 65$ K for A and 270 $\pm 100$ K for
B.  For both components, the non-thermal velocity contribution to the
linewidths is $\sim 1.5$ km s$^{-1}$.
We note that these temperatures apply to the atomic species, 
and  not necessarily  to  CO and CH because these molecules appear to
be kinematically-decoupled from the atomic species.

\subsubsection{Warm gas}

Components 2 to 4 belong to the second type of interstellar absorption lines, and T is estimated using the 
UV-line fitting procedure. 
Component 2 has a temperature typical of warm diffuse gas ($7\,000$--$10\,000\,{\rm K}$),
while in Component 3 and 4 the thermal broadening is clearly dominant,
with temperatures of $13\,000\pm3\,000\,{\rm K}$ for component 3, and about
$20\,000\,{\rm K}$  for component 4 as derived from weak and moderate lines.  In
the latter, a more precise determination of T is
impossible because gas of different temperatures appear to be present at this
velocity.
\subsubsection{ Hot gas}
Indeed the strong Mg~{\sc ii} lines show the presence of additional  high
temperature gas at approximately $200\,000\,{\rm K}$, moving at a velocity close to
that of component 4. The presence of high temperatures 
is well-illustrated  by the clear difference between the blue wings of the
Mg~{\sc ii} profile and  the N~{\sc i} profile in
Figure~\ref{fig:ions}. While the
relatively steep blue wing of the N~{\sc i} line can be fitted with a
temperature close to $20\,000\,{\rm K}$ for component 4, the extended
blue wing in Mg~{\sc ii} requires the addition of a hot
component at the same velocity, of a temperature $\sim 200\,000\,{\rm
K}$, and a column density 
 about 10\% of that of the  total column density of  component 4.  If this extended
blue wing was a damping wing due to a component with very high column
density, this component would create a significative absorption in the
faint Mg~{\sc ii} line at 1240  \AA, at least as strong as that of component 1. However, no
additional component is detected significantly  in Mg~{\sc ii} $\lambda$1240\AA. The
blue wing can therefore only be due to high-temperature, low-column
density gas.

High temperatures are also detected for the excited species.
The presence of hot gas is traced by the large width of the Si~{\sc ii}* lines compared 
to the width of other lines of moderate
strength. The FWHM of the Si~{\sc ii}* line, shown in
Figure~\ref{fig:ions}, is $\sim18$~${\rm {km\, s}^{-1}}$, corresponding
to $\sim200\,000\,{\rm K}$. In the C~{\sc ii}* line, the blue
wing is extended, up to LSR velocities of $-50\,{\rm {km\, s}^{-1}}$.
This may correspond to the presence of very
high temperature gas at a velocity close to that of component~4, as in the case of Mg~{\sc ii}.

The high-ionization species Si~{\sc iii}, Si~{\sc iv} and C~{\sc iv} have been detected in
the spectra at the velocity of component~4. They are broad and the
temperature implied by their width is  {        at least  $200\,000\,$ K. 
A temperature close to $10^{6}$ K, is in fact consistent with the observations (see figure 4 for an 
illustration), although the large width of the lines could  be due to the presence of several 
components at a temperature of  a few
$10^{5}$ K.}
OVI is
unfortunately not available in absorption in the FUSE spectra, because
the stellar continuum is too weak at these wavelengths.

\subsection{Summary of the line of sight}

The description of the line of sight that emerges from
the direct analysis of the spectral data 
is complex. The bulk of the gas is at low LSR velocity. Neutral low-potential 
atoms  and most of the  H~{\sc i} are distributed over two components separated 
by $\sim 3 \kms $(called A and B) 
that appear in these species to have temperatures of at most a few hundred K,
and are  associated with CH$^{+}$. The molecules CO and CH however,
are within a single component of similar b-value ($b \sim 1.5 \kms $) at a 
velocity intermediate between that of A and B.

Other components exist at high negative velocities with respect to the
bulk of the gas. Negative-velocity gas is seen in the absorption 
lines of  ions and  atoms with ionization potentials higher than that of H~{\sc i}, as well as in 
emission in H~{\sc i}.
The temperature of these components ranges from  typical temperatures of diffuse atomic gas 
($\sim 10\,000 \,$K) to somewhat higher temperatures (comparable to $20\,000\, $K), and 
increases with velocity offset from the bulk of the gas, in addition to the gas excitation state.
The highest velocity component, while evident for  moderately-ionized species, of temperature 
approximately  $20\,000\, $K,   is associated with highly-excited, highly-ionized  hot gas of
temperatures up to several $10^{5}$K.

\section{The cool gas and its traces of warm gas}\label{sec:model}

The bulk of the gas detected along the line of sight, is cool diffuse molecular gas at low velocities.
We comment on its properties within the framework of diffuse clouds studies. 

Diffuse molecular clouds are often modeled as extended, homogeneous, low-density structures
with properties regulated by the progressive
extinction of the UV field by dust. 
In a companion paper (Nehm\'e et al. 2007), we show that 
most observations of the cold, low-velocity gas, namely the H$_2$ abundance and
ortho-to-para ratio,
the CO abundance, the C~{\sc i} abundance and excitation, and the 158$\mu $m C~{\sc ii} 
line emission fit with model computations for a 
diffuse molecular cloud of a gas density of $80\,$cm$^{-3}$, 
illuminated  by a UV radiation field  close to the mean Solar Neighborhood value. 
The gas  temperature, derived from the balance between cooling and heating processes, 
ranges from 60 to 80~K. Combining the gas density and gas temperature we calculate
a thermal pressure, $P/k \sim 5000 {\rm cm}^{-3}{\rm K}$, which is within the range of values 
measured by Jenkins (2002), for clouds within the local Bubble. 
While the physical conditions inferred from the
model probably apply correctly to  the bulk of the gas, we hereafter discuss 
two points that show the limitations of the model's simplicity. 

The first point is the apparent kinematic separation of molecular and
atomic species in the cool gas, with the distribution of atomic species and CH$^{+}$  over two velocity components 
A and B, and the molecules CO and CH observed at a single intermediate value of velocity. 
As noted in Section~\ref{sub:structure}, we believe
that these components are related and therefore treat them as a single cloud in the modeling.
However, the observed velocity structure of the cloud cannot be accounted for by the photon-induced 
chemistry model 
where most of the C~{\sc i}, CO and H$_2$ are predicted to occupy the same volume (see
figures 3 and 4 in Nehm\'e et al. 2007).

The second point  is the existence of traces of warm molecular gas, as indicated by 
a CH$^+$ absorption line, and a population of  $\rm J > 2$  H$_2$ rotational 
levels that exceed model values by almost two orders of magnitude. 
Falgarone et al. (2005) determined 
an average Galactic-fraction of warm H$_2$,
in cool diffuse gas, expressed as the ratio of the
column density of H$_2$ molecules in levels $J>2$ (or $N({\rm H}_2^*)$
per magnitude of gas
sampled: $N({\rm H}_2^*) /A_V= 4 \times 10^{17}\,$
cm$^{-2}$. The warm H$_2$ predicted  along the line of sight to HD~102065, with 
this average Galactic fraction,
would then equal $N({\rm H}_2^*)=2.7 \times
10^{17}\,$cm$^{-2}$,  compared to the range 1-3 $\times 10^{17}\,$cm$^{-2}$
given in Table 1. 
The ratio of the CH$^+$ to the total hydrogen column densities
towards HD~102065 $N({\rm CH^+})/N_H = 1.2 \times 10^{-7}$, 
is comparable to the values calculated in
previous observations by Crane, Lambert \& Scheffer (1995) and 
Gredel (1997). 
The formation of CH$^+$ in diffuse gas
involves the endothermic reaction C$^+$ + H$_2 \rightarrow$
CH$^+$ + H ($\Delta E/k = 4640$ K), 
and has to be triggered by deposition of supra-thermal energy
either in MHD shocks (Flower \& Pineau des For\^ets 1998), or 
in large velocity shears (Joulain et al. 1998).

The chemical patterns triggered by the local deposition of non-thermal
energy involve several other endothermic reactions, including CH
destruction, which produces neutral carbon. 
According to the models explored in Joulain et al. (1998), 
the column density of C~{\sc i} produced in the warm chemistry $N({\rm
C_w})$, scales with that of CH$^+$ as $N({\rm C}_w) \approx 3 N({\rm CH^+})$
or $N({\rm C}_w) \approx  3\times 10^{13}$ cm$^{-2}$.
This corresponds to less than 10\% of the total C~{\sc i} column density
 $N({\rm C})=5.7 \times 10^{14}$ cm$^{-2}$ detected 
in components A and B (see Table~\ref{tab:abscolumn}).
The domain of parameters influencing the out-of-equilibrium
chemistry is difficult to  explore, and this is only an estimate
provided by existing models. 
The production of CO is enhanced by the warm chemistry patterns,
due to enhanced production in the warm gas  of 
OH, HCO$^+$ and CH$_3^+$,  all chemical precursors of CO.
It is possible that the observed heterogeneity  of CO abundances 
discussed above, originates in such processes.
\section{HD~102065 gas components within the local bubble context}\label{sec:context}

In this section, we place the HD~102065 observations in a 
broader context by relating them to the 
interaction between the nearby interstellar medium, and the Sco-Cen 
OB association, the Local  and the Loop I Bubbles.

Based on an extensive study of colour excess versus distance for a large sample of stars 
within $\rm 294^\circ <l< 308^\circ$ and $-20^\circ< b <5^\circ$, Corradi et al. (1997) 
concluded that the local density of matter is low until an extended interstellar dust feature 
at $150\pm 30 \,$pc from the Sun associated with the Chamaeleon, Musca and Southern Coalsack dark clouds.
The same authors complemented their investigation with NaI absorption spectra. Absorption lines
at negative velocities are detected towards many stars including the stars closest to the Sun 
(Corradi et al. 2004), which is indicative of a low column density ($\rm N_H \sim$ 
a few  $\rm 10^{19} \, cm^{-2}$), nearby ($d<60\, $pc) sheet of gas, outflowing from the  
Scorpius-Centaurus stars with mean radial velocity $\rm V_{LSR} = -7 \kms $. 
The presence of such a sheet was first proposed by Egger and Aschenbach (1995) to account for
the soft X-ray shadow seen in the ROSAT All Sky Survey maps.  

In the northern sky, the Loop I shell extends up to North Polar radio spur. 
In the southern sky, a broad H~{\sc i} filament 
at $b=-35^\circ $ and $\rm V_{LSR}\sim -20 \kms $
marks the outer extension of gas swept up by the 10-15~Myr old LCC and UCL  subgroups (de Geus 1992).

\begin{figure}[!htbp]
\begin{center}
\includegraphics[width=1.1\columnwidth]{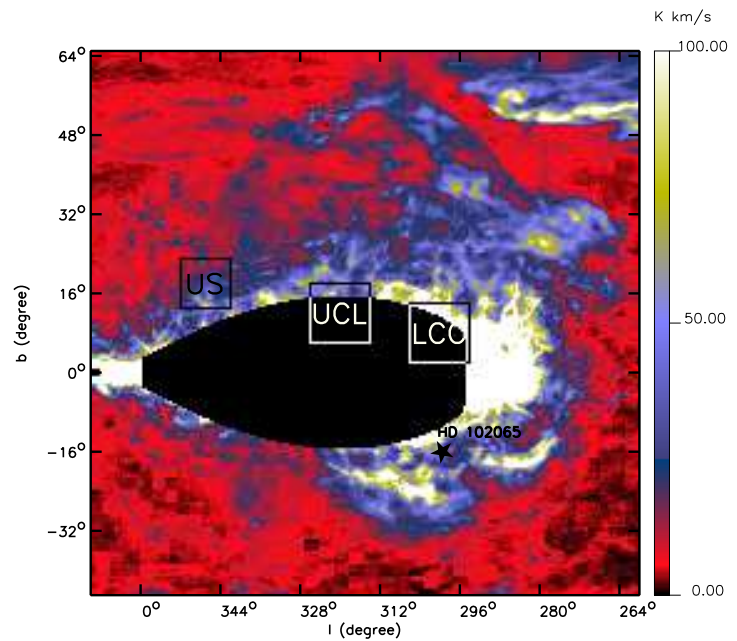}
\end{center}
\caption{H~{\sc i} emission at negative velocities $ \rm -50<V_{LSR}<-20 \kms $ in direction
of the Sco-Cen association.  H~{\sc i} observations from  Bajaja et al. (2005) and Kalberla et al. (2005). 
The position of HD 102065 and the US (Upper-Scorpius), UCL (Upper Centaurus
Lupus) and LCC (Lower Centaurus Crux) sub-groups of the OB association
are marked. The black area indicates the region where the selected range of 
negative velocities overlaps  with the
Galactic rotational velocities of distant gas.  The electronic version is in color.
\label{fig:HIneg} }
\end{figure}

In the Argentine H~{\sc i} survey (Kalberla et al. 2005 and Bajaja et al. 2005), 
H~{\sc i} gas at velocities $ \rm V_{LSR}< -20 \kms $, down to 
$-50  \kms $, is visible 
over $60^\circ$ in longitude about the Sco-Cen sub-groups (see Figure~\ref{fig:HIneg}).
The selection in radial velocity focusses on gas streaming towards the Sun at the center of
Loop I. We note also that much of the negative velocity gas close to the Upper-Scorpius (US) group that
still contains O stars must be photo-ionized and thus not seen in H~{\sc i}.  
The H~{\sc i} survey data show that negative velocity gas  
is clumped and scattered over a wide range of negative velocities, which   
cannot be accounted for within the 
simple picture of an expanding shell about a localized group of massive stars. We  propose instead that 
we are looking at dispersed fragments of the parent cloud of the LCC and UCL stars accelerated
to a range of velocities by supernovae blast waves. 
Since the stellar groups have moved with respect to the Sun (de Zeeuw et al. 1999), 
the explosions have occured at different locations and
distances from the Sun fragmenting and spreading the stars' parent cloud between the
Local and Loop I bubbles. This view accounts for the stream of 
negative-velocity gas observed close to the Sun, including the local cloud about the Sun,  
detected at negative velocities,
streaming from the general direction of the Sco-Cen association (Ferlet 1999).
Our view of the relation between the nearby interstellar medium and the Sco-Cen stars, 
the Local and Loop I bubbles  builds on the original scenario proposed 
by Breitschwerdt et al. (2000). However, the observations suggest a more complex picture
where there is not a continuous sheet of swept-up gas separating the two bubbles, 
but instead clouds of cold and warm gas spread out between the two bubbles. 

We believe that the diffuse molecular cloud in the foreground of HD~102065, is such a cloud. 
As discussed by
Mizuno et al. (2001), there is strong evidence that Dcld 300.2-16.9 {        (the Blue Cloud)} 
is not part of  the Chamaeleon molecular clouds complex.  
A distance of $70\pm15\, $pc 
from the Sun, is inferred from the parallax of the T-Tauri star
T~Cha, which coincides in position and velocity (Covino et al. 1997) 
with the cloud CO emission. This places the cloud at a distance of 
$\sim 30$ pc from the center of the LCC sub-group (de Geus 1992). Its cometary
structure, with a tail extending in the
opposite direction to that of the LCC stars, suggests that it could have been shaped 
by a supernova explosion from one of the stars. 
With a total mass of $\rm \sim 100 \, {\rm M_\odot}$ (including the tail, 
as calculated by Boulanger et al. (1998) and corrected assuming  a distance of $70\,$pc), the cloud is
too massive to 
have been  globally accelerated by a distant explosion 
but its structure suggests that  gas ablated from the surface by the blast wave is streaming outwards.
In the H~{\sc i} data, the low velocity components A and B are connected
to the cloud head but extend in different directions along the cloud tail
(Figure~\ref{Cham12_HI}). This may imply that they represent both sides
 of the gas flow, that they encompass
the cloud core, 
and that the difference in radial velocity is the result
of a difference in angle between the streaming direction, and the line of sight. 
{        At the surface of the gas flow, the  CO and CH molecules are photodissociated, explaining 
why they are not detected in components A and B, but instead in a single component C, which would
correspond to the internal part of the gas flow.}
To account for the $3 \kms $ velocity difference between components A and B
within a reasonable angle difference ($\sim 45^\circ$), the gas streaming velocity
must be $\sim 10  \kms $. 
The tail length $\sim 2^\circ$ ($2.4\, $pc for a distance of $70\, $pc)
translates into a time elapsed since the supernova explosion 
of $\sim 2-3 \times 10^5\, $yr. 
The HD~102065 observations were originally motivated by the strong mid-IR 
emission of the foreground cloud
interpreted as an unusually high abundance of small dust grains \citep{Boulanger90}. 
This distinct characteristic may be related to the position of the cloud within the Local Bubble, and 
its plausible interaction with a supernova blast wave. 

The interaction between interstellar clouds and shock waves travelling in their surrounding medium, 
has been investigated using numerical simulations. 
Images presented in Nakamura et al. (2006) help to picture the complexity of the
interaction that ablates matter from the cloud and generates a turbulent flow streaming away from 
the cloud. The fact that the 
C~{\sc iv} and Si~{\sc iv} absorption lines peak at negative velocities qualitatively 
agrees with this picture where one expects the
low density gas in the turbulent mixing layer to be entrained by the flow of hot gas. The 
discrepancy between gas temperature and ionization state evident 
{        for some of the } Mg~{\sc ii}, Si~{\sc ii}$^{*}$ and C~{\sc ii}$^{*}$  
 observed at a temperature $2\times 10^5\,$K,  implies that dynamical mixing of 
hot and warm gas, on timescales shorter than that necessary to reach ionization equilibrium, is taking place.

\section{Conclusion}
\label{sec:Conclusions}

High resolution spectroscopic observations are presented and used to characterize the 
physical state and velocity structure of the multiphase interstellar medium observed
towards the nearby star HD~102065. These spectroscopic observations provide detailed
information on the kinematics and physical state of the gas, along the line of sight.

\begin{itemize}

\item
Gas is observed over a 
wide range of velocities, with cold molecular gas at low velocities, and warmer and lower density gas  
at negative velocities. Most of the hydrogen column density is in cold diffuse molecular gas at 
low LSR velocities. 

\item
The spectra  show three distinct components 
at negative velocities as low as $\rm V_{LSR} = -20 \kms $. 
The excitation and temperature of the species detected in the negative 
velocity components, increase with the velocity separation from that of the
diffuse cloud. H~{\sc i} emission maps show that the 
negative velocity gas extends over $\rm \sim 200 \, {\rm deg}^2$. It 
is fragmented and spreads over velocities 
down to $-50 \kms $. 

\item
A striking result of the data analysis is the detection of gas out of equilibrium
in both the low and the negative velocity gas.
At low velocities, we find that the cold molecular gas is mixed 
with traces of warmer molecular gas  where  
the observed CH$^+$ must be formed and which is also traced by the observed excess 
population of H$_2$ in the $J>2$ levels.  
At negative velocities, gas in a low ionization state (Mg~{\sc ii}, Si~{\sc ii}$^{*}$ and C~{\sc ii}$^{*}$) is observed far
out of ionization equilibrium at a  temperature of $2\times 10^5$K. 

\end{itemize}

We have set the observational results within the scenario  
proposed by Maiz-Apellaniz (2001) and Bergh{\"o}fer and Breitschwerdt (2002)
that connects the origin of the Local Bubble to supernovae explosions from early star
members of the oldest Sco-Cen groups.

\begin{itemize}
\item
The gas at 
low LSR velocities is observed to be associated with the Blue
Cloud (Dcld 300.2-16.9) located within the local bubble at a distance of $\sim 70$pc from the Sun. 
Its cometary shape with a tail pointing away from the 
LCC sub-group of the Sco-Cen  association $\sim 30$pc away, strongly suggests that it has been shaped by 
the interaction with a recent ($2-3 \times 10^5\, $yr) 
supernova explosion from one of the LCC stars. The line of sight to HD~102065 
goes across the cloud tail. The velocity structure of the low velocity gas may be understood as the 
two sides of a flow of gas ablated from the cloud head Dcld 300.2-16.9. 
The large abundance of small dust particles inferred from the IRAS colors 
of the Blue Cloud tail, possibly results from  
the cloud interaction with the supernova blast wave.

\item
The negative velocity gas must have been accelerated by a supernova
explosion that  occurred in the Sco-Cen OB association. 
Our interpretation of the Blue Cloud shape would imply 
that it is physically connected with the lower velocity gas.  
The column density of Si~{\sc ii}$^*$ at $-20 \kms $ might be a signature
of an ongoing shock.
\end{itemize}
A complete diversity of  diffuse interstellar-medium components, is
present along the line of sight.
Our analysis of both low and negative velocity, hot, warm and cold gas 
components, stresses the complex dynamical interplay between the
interstellar medium phases, in agreement with the predictions of  
numerical simulations (Audit and Hennebelle  2005,
de Avillez and  Breitschwerdt  2005).  
We started this work by considering an apparently unusual cloud, but now believe 
that the questions we are now asking, have significance for the general study of diffuse, 
interstellar matter in the Galaxy.

\begin{acknowledgements} 

This work has been done using the profile fitting procedure Owens.f
developed by M. Lemoine and the FUSE French team
(http://www2.iap.fr/Fuse/qoutils.html).  We thank E. Arnal, M. Eidelsberg and
M. Rubio for providing data prior to publication and
F. Viallefond for his help in using Gipsy software \citep{VT(2001)}.
FB acknowldeges fruitful discussions with M. de Avillez and R. Lallement.

\end{acknowledgements}

\end{document}